\documentclass[twocolumn]{aastex62}

\shorttitle{Superclusters in the Sloan Digital Sky Survey}
\shortauthors{Sankhyayan et al.}

\begin{document}

\title{Identification of Superclusters and their Properties in the Sloan Digital Sky Survey Using WHL Cluster Catalog}

\correspondingauthor{Shishir Sankhyayan}
\email{shishir.sankhyayan@ut.ee, shishir9898@gmail.com}

\author[0000-0003-2601-2707]{Shishir Sankhyayan}
\affil{Tartu Observatory, University of Tartu, Observatooriumi~1, 61602 T\~oravere, Estonia}

\author[0000-0002-2922-2884]{Joydeep Bagchi}
\affiliation{Department of Physics and Electronics, CHRIST (Deemed to be University), Bengaluru-560029, India}

\author[0000-0002-5249-7018]{Elmo Tempel}
\affiliation{Tartu Observatory, University of Tartu, Observatooriumi~1, 61602 T\~oravere, Estonia}
\affiliation{Estonian Academy of Sciences, Kohtu 6, 10130 Tallinn, Estonia}

\author[0000-0002-2986-2371]{Surhud More}
\affiliation{The Inter-University Centre for Astronomy and Astrophysics (IUCAA), S.P. Pune University Campus, Post Bag 4, Pune 411007, India}
\affiliation{Kavli Institute for the Physics and Mathematics of the Universe (IPMU), 5-1-5 Kashiwanoha, Kashiwashi, Chiba 277-8583, Japan}

\author[0000-0003-3722-8239]{Maret Einasto}
\affiliation{Tartu Observatory, University of Tartu, Observatooriumi~1, 61602 T\~oravere, Estonia}

\author[0000-0001-9212-3574]{Pratik Dabhade}
\affiliation{Instituto de Astrof\' isica de Canarias, Calle V\' ia L\'actea, s/n, E-38205, La Laguna, Tenerife, Spain}
\affiliation{Universidad de La Laguna (ULL), Departamento de Astrofisica,
E-38206, Tenerife, Spain}

\author[0000-0002-4864-4046]{Somak Raychaudhury}
\affiliation{Department of Physics, Ashoka University, Sonipat, Haryana 131029, India}
\affiliation{The Inter-University Centre for Astronomy and Astrophysics (IUCAA), S.P. Pune University Campus, Post Bag 4, Pune 411007, India}

\author[0000-0001-7141-7311]{Ramana Athreya}
\affiliation{Indian Institute of Science Education and Research (IISER), Dr. Homi Bhabha Road, Pashan, Pune 411008, India}

\author[0000-0002-1568-0227]{Pekka Heinämäki}
\affiliation{Tuorla Observatory, Department of Physics and Astronomy, University of Turku, Finland}

\begin{abstract}
Superclusters are the largest massive structures in the cosmic web on tens to hundreds of megaparsecs (Mpc) scales. They are the largest assembly of galaxy clusters in the Universe. Apart from a few detailed studies of such structures, their evolutionary mechanism is still an open question. In order to address and answer the relevant questions, a statistically significant, large catalog of superclusters covering a wide range of redshifts and sky areas is essential. Here, we present a large catalog of 662 superclusters identified using a modified {\it Friends of Friends} algorithm applied on the WHL (Wen-Han-Liu) cluster catalog within a redshift range of $0.05 \le z \le 0.42$.
We name the most massive supercluster at $z \sim 0.25$ as \textit{Einasto Supercluster}.
We find that the median mass of superclusters is $\sim 5.8 \times 10^{15}$ M$_{\odot}$ and median size $\sim 65$ Mpc. We find that the supercluster environment slightly affects the growth of clusters. We compare the properties of the observed superclusters with the mock superclusters extracted from the Horizon Run 4 cosmological simulation. The properties of superclusters in mocks and observations are in broad agreement.
We find that the density contrast of a supercluster is correlated with its maximum extent with a power law index, $\alpha \sim -2$. The phase-space distribution of mock superclusters shows that, on average, $\sim 90\%$ part of a supercluster has a gravitational influence on its constituents.
We also show mock halos' average number density and peculiar velocity profiles in and around the superclusters.
\end{abstract}

\keywords{Superclusters (1657) --- Galaxy clusters (584) --- Catalogs (205) --- Large-scale structure of the universe (902)}

\section{Introduction}
\label{sec:intro}
The intricate network of the distribution of galaxies and matter in the Universe is called the Cosmic Web \citep{Bond96}. The details of the cosmic web help in our understanding of the cosmological models
governing the evolution of structures in the Universe. The main components of the cosmic web are clusters, two-dimensional walls, one-dimensional filaments, and under-dense regions called voids.

In the cosmic web, there exist large coherent regions ($\sim$ 10 -- 100 Mpc) that are larger than the dimensions of galaxy clusters which span about a few Mpc. Moreover, these regions are extremely massive and contain several groups and clusters of galaxies apart from the galaxies and intergalactic dark and baryonic matter in between the clusters. Cumulatively, these regions are called superclusters. Superclusters are not as abundant as galaxies or even as groups and clusters, but they are known to affect the evolution of galaxies  within them \citep[e.g.][]{EinastoM07,EinastoM11_SGW,EinastoM14,Lietzen16p}. It is still an open question how these giant structures form and evolve
in the cosmic web, and answering them requires detailed studies involving observations and simulations. Although this is a growing field of research that is getting enriched  with  the availability of more  data, a single widely accepted definition of a supercluster does not exist as yet.
Superclusters are the coherent regions in the cosmic web that have been defined in different ways in the literature. The two extreme definitions of a supercluster are based on the largest bound structures in the Universe and the largest regions with converging peculiar velocity field flows.
The superclusters in the literature can be broadly  divided into three categories:
(1) the gravitationally bound regions, (2) the unbound over-dense regions in the Universe, and (3) the converging peculiar velocity field regions. According to definition (1), 
they have been defined as the largest over-dense regions with sufficiently high matter density to overcome the global expansion of the Universe. They would eventually collapse and form gravitationally bound systems \citep{Dunner06,Araya-Melo09,Luparello11,Chon15}.
According to definition (3),
using the peculiar velocity field, they have been defined as regions in which, on average, peculiar velocities of galaxies converge \citep{Tully14,Pomarede15,Dupuy19,Pomarede20}.
These superclusters form regions of dynamical influence where they act as great attractors which grow by the inflow of matter from lower-density regions \citep{Tully14}.
\citet{Einasto19} and \citet{Dupuy19} showed that the whole cosmic web can be divided into regions of dynamical influence or the basins of attraction.
This means that type (3) superclusters contain some parts of the under-dense void regions surrounding them.
Observationally, Laniakea -- our home supercluster \citep{Tully14}, has been identified using the rich observational data of the peculiar velocities of the nearby galaxies. But, such data is  still not
available for distant galaxies; therefore, the velocity flow field method of estimating regions of convergence can not be derived for them. However, these converging peculiar velocity field regions (or the `basins of attraction') can be identified in simulations.
It is often seen that the supercluster definition (1)
of gravitationally bound systems picks up the central regions of superclusters defined with other criteria (definitions (2) and (3)), for example, over-densities in the luminosity density field \citep[e.g.][]{EinastoM22}.
Definition (2) picks up the over-densities present in the matter density field. These regions do not contain any parts of under-dense void regions. The over-density threshold to consider in defining these types of superclusters is not set, and depending on the focus of the study, it varies in the literature \citep[see for example,][]{Einasto07,Liivamagi12,Lietzen16,Einasto19}.

Superclusters are traditionally identified using the friends-of-friends (FoF) algorithm. FoF is also used to find bound halos in simulations \citep{Knebe11}. FoF has been used in several supercluster finding works \citep[e.g.][]{EinastoM97,Chow-Martinez14,Bagchi17} in various forms using clusters as well as galaxies as their input data. Apart from FoF, there are other methods as well to extract superclusters. For example, by applying different density threshold cuts on the luminosity density field of the galaxy distribution \citep{Einasto07,Liivamagi12, Lietzen16} and watershed method by applying it on the number density field constructed by Voronoi tessellation \citep{Neyrinck08,Nadathur16b}.

Recent studies show that a supercluster's size, mass, and luminosity show some evolution with redshift. In this context, luminosity represents the total luminosity of all galaxies within the supercluster. Size refers to the maximum comoving extent of the supercluster, and the mass indicates the total amount of baryonic and dark matter contained within the supercluster.
As the luminosity and mass increase with time, the overall size decreases \citep{Araya-Melo09,Einasto19,Einasto21}. The evolution of galaxies inside a supercluster is affected by its host supercluster's environment \citep{Lietzen2012, Seth2020, Alfaro2022}. Also, the evolution of clusters and groups could be governed by the environment of superclusters in which they reside. Extremely large superclusters may have a linear growth on average, but on a megaparsec (Mpc) scale, galaxies and clusters have a non-linear growth within them. Overall, massive and large superclusters are excellent targets to study the diversity of environments they offer to their resident galaxies. Multi-wavelength observations of such objects will aid our understanding of their growth and evolution. Ongoing and upcoming deep weak lensing surveys would immensely help in estimating the masses of these huge overdensities.

In the $\Lambda$ Cold Dark Matter ($\Lambda$CDM ) model of the Universe, the overall growth of structures  slows down below redshift $z \approx 0.5$ due to the influence of dark energy \citep{Frieman08,Einasto21skew}. However, the highly overdense structures are still able to grow. In a spherical collapse model, structures that are at the turnaround stage at the present epoch, with a density contrast of $\sim 13.1$, form bound systems. Structures with present-day density contrast of $\sim 8.73$ will reach the turnaround stage in the future \citep{EinastoM2020}. \citet{EinastoM21}, using the spherical collapse model, suggested that structures with a present density  contrast of $\approx 30$ (rich clusters and their regions of influence in supercluster cores) have passed turnaround and started to collapse at redshifts $z \approx 0.3 - 0.4$. Therefore, a supercluster may or may not collapse depending on the average density contrast of the entire supercluster. For example, 
\citet{EinastoM22} show that the BOSS Great Wall fragments into different structures in physical space, and these fragmented structures will collapse individually in the future.

In order to enhance our understanding of superclusters and their properties, it is important to identify these structures over a broad range of redshifts and a wide area of the sky. Moreover, their dynamical behavior can be explored using simulations. This paper presents 662 superclusters identified within a redshift range of $0.05 \leq z \leq 0.42$, covering a sky area of $\sim$ 14,000 deg$^2$. The properties of these superclusters are presented and compared with simulations.
The layout of the paper is as follows.
In Section~\ref{sec:data}, we present the observational data and mocks created through simulation. In Section~\ref{sec:method}, we present our analysis methods. In Section~\ref{sec:prop_super}, we define the properties of the superclusters. Section~\ref{sec:result_all} presents the results of our analysis. The properties of the identified superclusters highlighting the most massive superclusters (e.g., Corona-Borealis \citep{EinastoM21}, Sloan Great Wall \citep{Gott05}, and Saraswati superclusters \citep{Bagchi17}), a power law relation between density contrast and the size of the superclusters, and the effect of the supercluster environment on the cluster properties are presented in Section~\ref{sec:result}. The comparison of the properties of superclusters extracted from the observations and simulations and the phase-space distribution of the mock superclusters are presented in Section~\ref{sec:result_sim}. We conclude and discuss the future prospects in Section~\ref{sec:conc}, and Appendix~\ref{sec:top5} highlights the five most massive superclusters.

Throughout this paper, we have adopted the following cosmological parameters: $H_0 = 72$ km s$^{-1}$ Mpc$^{-1}$, $\Omega_m = 0.26$ and $\Omega_{\Lambda} = 0.74$.
R$_{\Delta c}$ is defined as the radius of a spherical region within which the matter density is $\Delta$ times the critical density ($\rho_c$) of the Universe. M$_{\Delta c}$ is the mass within this spherical region of radius R$_{\Delta c}$.
So, M$_{500c}$ denotes the mass within a spherical region of radius R$_{500c}$ where the mass density is 500 times the critical density of the Universe.

\section{Data}
\label{sec:data}
Here, we describe the observational data and simulation used in our analyses.
\subsection{Observational Data}
\citet{Bagchi17} used the spectroscopic sample of galaxies in the Sloan Digital Sky Survey (SDSS; \citealt{York00}) to identify the massive supercluster named Saraswati. We have been motivated by the large excess of clusters and groups \citep{Wen12} found in the Saraswati supercluster region \citep{Bagchi17}.
As a result, we decided to search for other superclusters in the SDSS using the same cluster catalog.

To achieve this, we use the group and cluster catalog of \citet{Wen15} (hereafter WH15\footnote{\url{https://cdsarc.cds.unistra.fr/viz-bin/cat/J/ApJ/807/178}}), which is an updated version of \citet{Wen12} (hereafter WHL12\footnote{\url{https://cdsarc.cds.unistra.fr/viz-bin/cat/J/ApJS/199/34}}). The WHL12 catalog consists of 132,684 groups and clusters identified from the photometric redshift data of galaxies within the redshift range $0.05 \leq z \leq 0.8$ from the SDSS-III\footnote{\url{https://www.sdss3.org}}. Galaxies with deblending problems and saturated objects were excluded. With the release of SDSS Data Release 12 \citep[DR12;][]{Alam15}, new spectroscopic redshifts of galaxies became available, leading to the addition of 25,419 rich clusters (primarily at high redshifts, $z > 0.4$) in the WH15 catalog.

\begin{figure*}
\centering
\includegraphics[width=\textwidth]{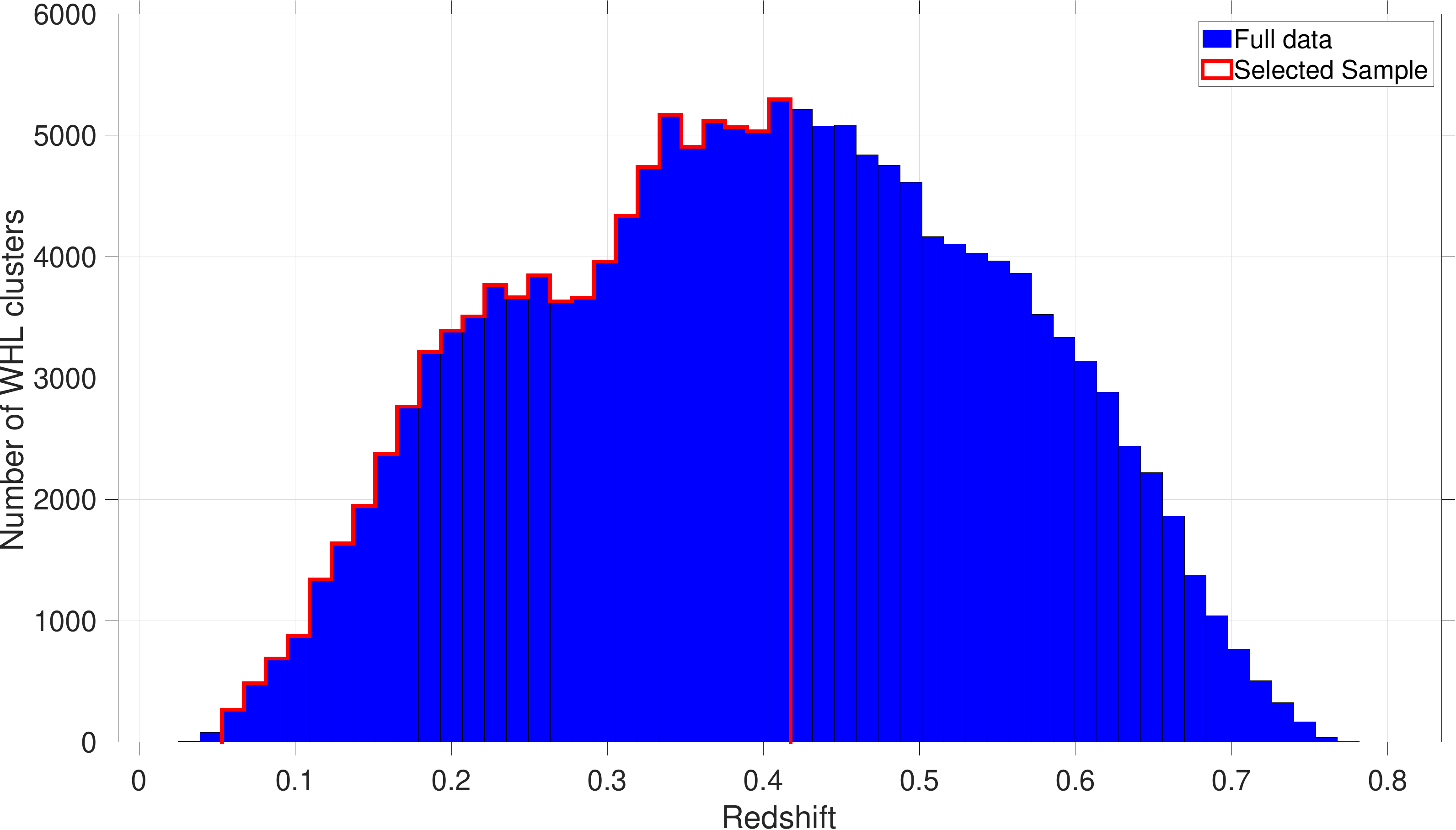}
\caption{Blue histogram represents the redshift distribution of the entire sample of 158,103 WHL clusters, and red represents the selected sample of 85,686 clusters within the redshift range of $0.05 \leq z \leq 0.42$.}
\label{fig:WHLredshiftdist}
\end{figure*}

In this paper, we utilize the spectroscopic redshifts available in DR12 for 89\% of the groups and clusters listed in the WH15 catalog. For the remaining 11\% of groups and clusters, we rely on photometric redshifts.
Figure~\ref{fig:WHLredshiftdist} shows the redshift distribution of WH15 groups and clusters.
The small peak observed at $z \sim 0.22$ and the dip at $z \sim 0.28$ are attributed to the redshift distributions of the two major samples (LRG and LOWZ galaxy samples) comprised in the SDSS.
The LRG sample exhibits an increase beyond $z \sim 0.2$, while the LOWZ sample peaks at $z \sim 0.35$. The dataset covers a sky area of $\sim 14,000$ deg$^2$, as shown in Figure~\ref{fig:whl-hr4-sel-fn}.

The completeness of the data, defined as the detection rate of the injected mock clusters using the cluster detection algorithm, depends on the virial mass $M_{200c}$.
For clusters with mass $\rm M_{200c} \gtrsim 0.6 \times 10^{14}~M_{\odot}$, the completeness is $\sim 80\%$ \citep{Wen12}. For clusters with $\rm M_{200c} > 10^{14}~M_{\odot}$, it exceeds 95\%, and reaches 100\% for clusters with $\rm M_{200c} > 2 \times 10^{14}~M_{\odot}$ within the redshift range $0.05 \leq z \leq 0.42$. Consequently, we focus on the data within this redshift range, selecting 85,686 groups and clusters to extract superclusters (Figure~\ref{fig:WHLredshiftdist}).
The false detection rate of these clusters is $\lesssim 6\%$ \citep{Wen12}.
For detailed group and cluster catalog construction information, refer to \citet{Wen12} and \citet{Wen15}. Throughout the rest of the paper, we will refer to the group and cluster catalog of WH15 (updated WHL12 catalog) as the WHL cluster catalog. Table~\ref{tab:WHLclustcat} shows the sample of selected 85,686 groups and clusters within the redshift range $0.05 \leq z \leq 0.42$.

The halo masses in WH15 are derived from optical richness, which was cross-calibrated using clusters having masses from X-ray and Sunyaev-Zeldovich (SZ) measurements. The optical richness is not expected to have a one-to-one correlation with the halo mass but will instead be a useful proxy rather than being entirely perfect and a proxy that will involve an intrinsic scatter. Therefore, a selection of clusters that is volume limited by mass cannot be obtained in practice. In addition, the various flux limits of spectroscopic and photometric surveys will imply that the members of the clusters at higher redshift will be limited to brighter galaxies, which on average, will be hosted by more massive clusters.
In principle, such selection effects can be thoroughly included in the mocks, but such an implementation will require a cluster catalog that calibrates both the mean and the scatter of the richness mass relation accounting for the flux limit effects mentioned above. In the absence of such calibration, including the intrinsic scatter is beyond the scope of this paper but is something that we plan to address in the future.
In the current work, we treat incompleteness to be independent of mass and compensate for the incompleteness of low-mass clusters at higher redshift by increasing the linking length of the existing clusters. This allows the clusters to be linked at larger distances in inverse proportion to the cube root of their density (see Section~\ref{sec:method}) and ensures that the identification of superclusters is not affected by the incompleteness of the chosen sample. Additionally, we preferred to use all data available in the WHL catalog as applying a higher mass cut means that we have to discard some observational data. Instead,  we use all available observational data and model the incompletenesses in the data. This is the approach we have adopted here while constructing the mock catalog.

\setlength{\tabcolsep}{8pt}
\begin{table*}
\centering
\caption{
Properties of 85,686 groups and clusters extracted from the WHL cluster catalog within the redshift range $0.05 \leq z \leq 0.42$. Columns (1 -- 4) are taken from the WHL catalog, and columns (5 -- 7) have been computed in the current paper (see Sections~\ref{sec:method} and \ref{sec:prop_super}). Columns represent: (1) Cluster identifier, (2) Right Ascension of the brightest cluster galaxy (BCG from now on), (3) Declination of the BCG, (4) Redshift of the BCG, (5) $R_{200c}$ of the cluster, (6) $M_{200c}$ of the cluster, (7) Supercluster number to which the cluster belongs. SCl = 0 means the cluster is not a part of any supercluster.
\label{tab:WHLclustcat}}
\begin{tabular}{|c|c|c|c|c|c|c|}
\hline
\hline
\textbf{ID} & \textbf{RA} & \textbf{Decl.} & \textbf{z} & $\mathbf{R_{200c}}$ & $\mathbf{M_{200c}}$ & \textbf{SCl} \\
 & \textbf{(deg)} & \textbf{(deg)} &  & \textbf{(Mpc)} & \textbf{($10^{14}$ M$_{\odot}$)} &  \\
(1) & (2) & (3) & (4) & (5) & (6) & (7) \\
\hline
J000000.6+321233 & 0.00236 &  32.20925 & 0.1274 & 1.72 & 5.9478 & 0 \\   
J000002.3+051718 & 0.00957 &  5.28827  & 0.1694 & 0.94 & 1.9627 & 0 \\
J000003.5+314708 & 0.01475 &  31.78564 & 0.0916 & 0.94 & 1.6766 & 0 \\  
J000004.7+022826 & 0.01945 &  2.47386  & 0.4179 & 0.95 & 0.4769 & 0 \\  
J000006.0+152548 & 0.02482 &  15.42990 & 0.1731 & 1.13 & 1.5391 & 0 \\
\hline
\end{tabular}
\tablecomments{This table will be available in its entirety in a machine-readable format in the online version of the journal. A portion is shown here for guidance regarding its form and content. The full table will also be available on CDS.}
\end{table*}

\subsection{Simulation and mock clusters}
\label{sec:mock}
We use the Horizon Run 4 (HR4) cosmological N-body simulation \citep{Kim15} to create a mock WHL cluster catalog. It has $6300^3$ dark matter particles in a cubic box of size 3150 h$^{-1}$Mpc.
The particle mass is $m_p \simeq 9 \times 10^9$ M$_{\odot}$, and the minimum mass of halos with 30 member particles is $M_S \simeq 2.7 \times 10^{11}$ M$_{\odot}$. The HR4 simulation adopted the \textit{Wilkinson Microwave Anisotropy Probe} (WMAP) 5-year \citep{Dunkley2009} $\Lambda$CDM cosmology : $\Omega_{m,0} = 0.26$, $\Omega_{\Lambda,0} = 0.74$, and $H_0 = 100~h$ km s$^{-1}$ Mpc$^{-1}$, where $h = 0.72$.

In order to create a mock catalog, we require information on the position, mass, and peculiar velocity of halos. However, if we consider the snapshot of the simulation at redshift $z = 0$, it does not correspond to the observations that capture the state of objects at different cosmological times (redshifts). Essentially, a halo (or a cluster in observations) becomes visible to the observer when it enters the observer's past light cone (or the center of the simulation box). Therefore, we need to consider the state of a halo (its comoving position, mass, and peculiar velocity) at the time (redshift) when it enters the observer's past light cone. The HR4 simulation output products provide the necessary past light cone information on halos.
We use the past lightcone space dark matter halos in the redshift range $0 < z \le 0.4$. We displace the halos according to their peculiar velocities to get the positions of halos in the redshift space (same  as the observations).

\begin{figure*}
\centering
\includegraphics[trim={4cm 0 6cm 0},clip, width=\textwidth]{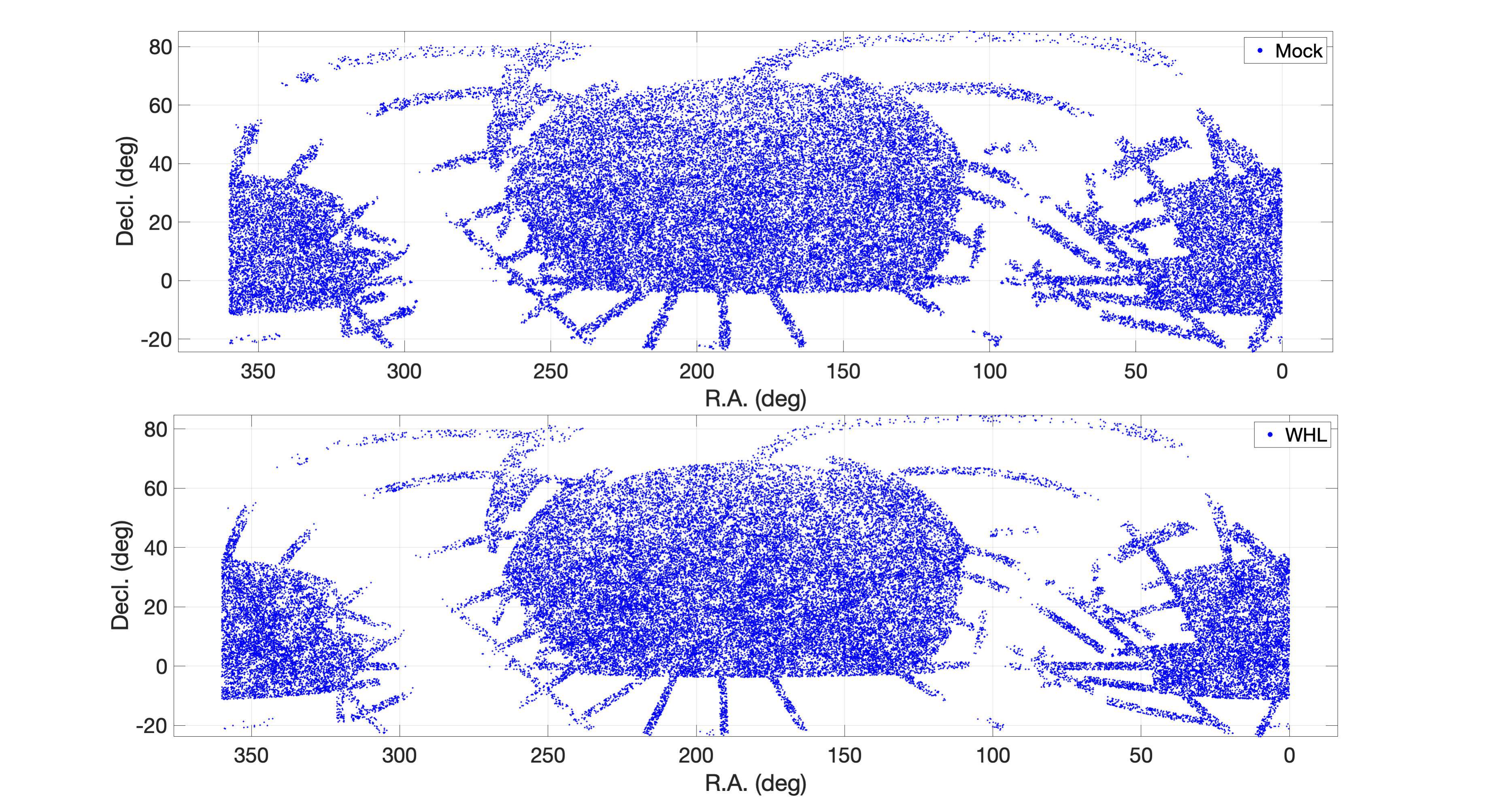}
\caption{Sky distribution of WHL clusters and HR4 halos within $0.05 \le z \le 0.366$. WHL clusters and HR4 halos are the observed and mock samples, respectively.}
\label{fig:whl-hr4-sel-fn}
\end{figure*}

\begin{figure*}
\centering
\gridline{\fig{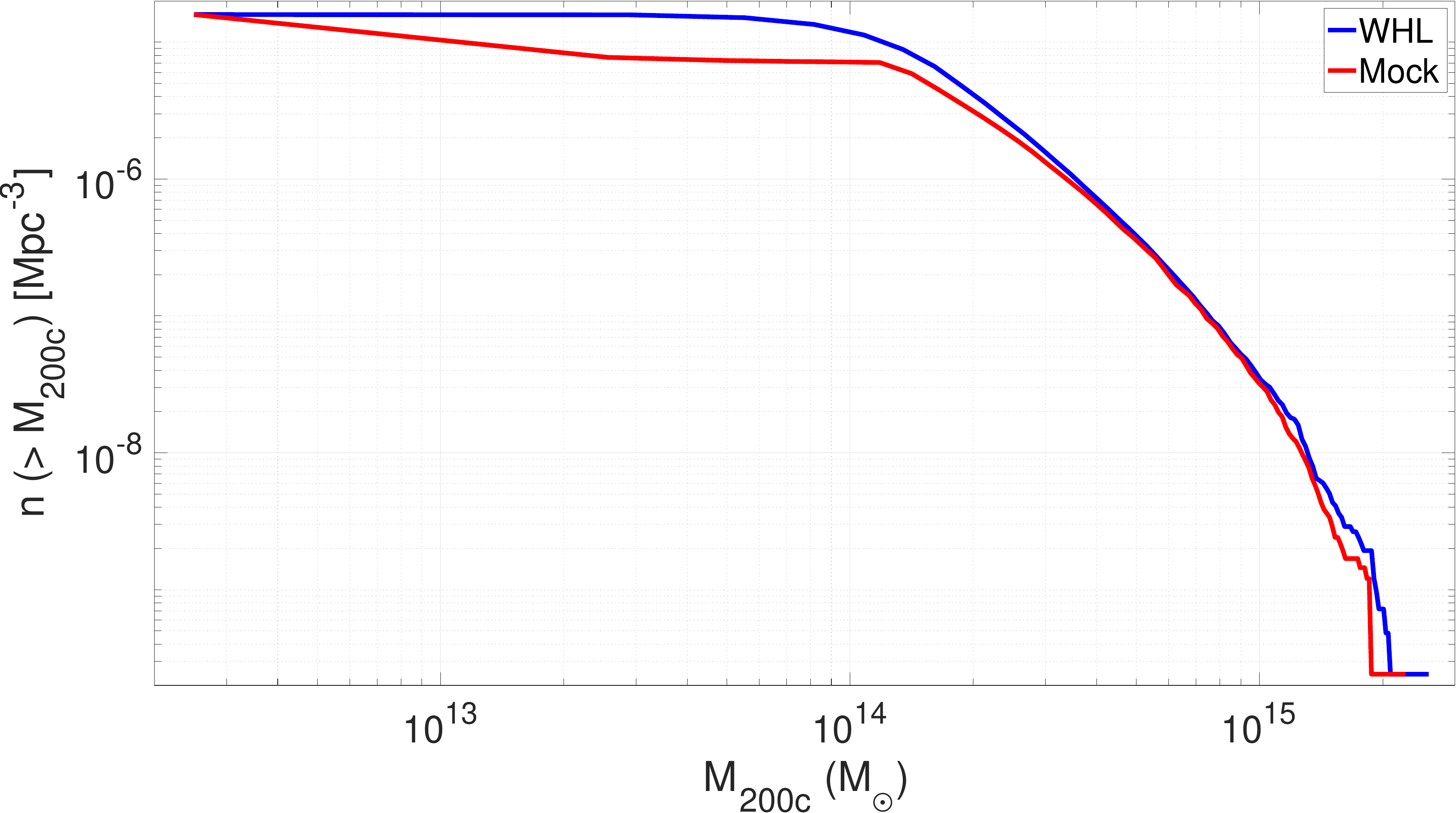}{0.475\textwidth}{(a)}
          \fig{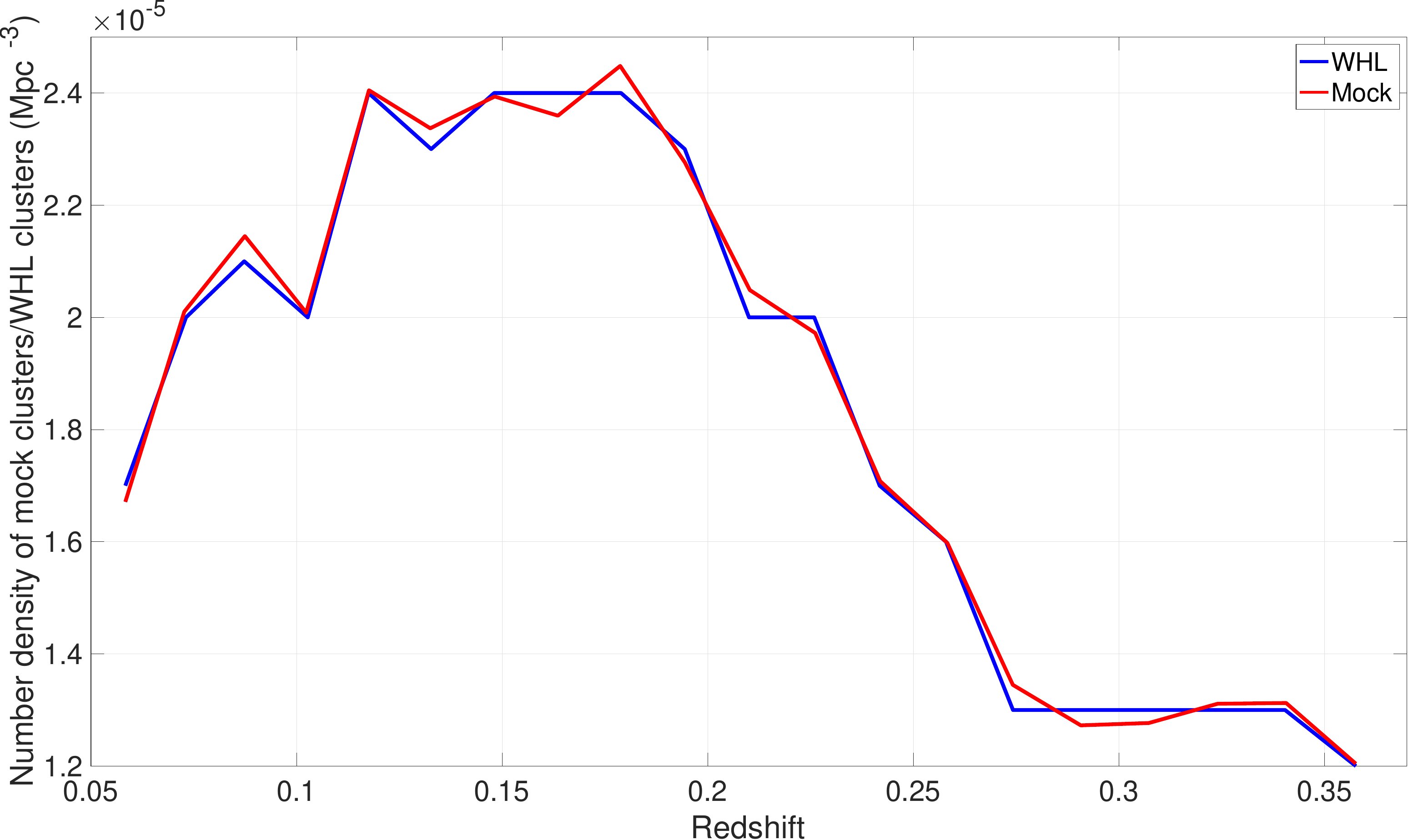}{0.5\textwidth}{(b)}}
\caption{(a) Mass functions of WHL clusters and mock clusters (HR4 halos) within $0.05 \le z \le 0.366$. $n$ is the number density of clusters/halos. (b) Number density as a function of the redshift of the WHL and mock clusters (HR4 halos) within $0.05 \le z \le 0.366$.}
\label{fig:mass-hmf}
\end{figure*}

The mock WHL cluster catalog is made by selecting the halos according to the selection functions of WHL cluster data. We matched the sky distribution, redshift distribution, and mass distribution of the WHL clusters for the extraction of halos from HR4 simulation (see Figure~\ref{fig:whl-hr4-sel-fn} and Figure~\ref{fig:mass-hmf}).
To make a sky mask for the WHL clusters' sky distribution, we use the Hierarchical Equal Area isoLatitude Pixelization (HealPix) scheme by \citet{Gorski05}. HEALPix divides the sky plane into equal area and shape-preserving pixels. We use $\rm N_{side} = 64$ and create the sky mask of the WHL catalog. A pixel is given an angular weight of 1 if it has at least one cluster in it and 0 if it does not contain any cluster. We then apply this mask to HR4 halos and extract all the halos on the sky pixels with a weight equal to 1. Next, we divide the redshift distribution into equal comoving distance bins. In each bin, halos are extracted to match the mass distribution and the number of WHL clusters in the corresponding bin.

Lastly, we find a small deficiency of large mass clusters in the HR4 simulation compared to WHL clusters. Similar deficiency with high luminosity galaxies (high mass end of galactic halos) is also seen in other studies \citep[see,][]{Trayford15,Tuominen21}.
To address this issue, when we encounter a deficiency of large mass clusters within a specific distance bin, we randomly select halos from the lower mass bin. However, we only choose halos that have not already been selected for the previous mass bin within that specific distance bin. This ensures that the number of mock clusters in each distance bin approaches the number of WHL clusters present in that particular distance bin. As a result, there is a small difference in the mass distribution between the WHL and mock clusters.

Additionally, to avoid the artificial change in the number density of halos at $z \sim 0.4$ due to the peculiar velocity of galaxies in the halos, we have limited the halos to $0.05 \le z \le 0.366$. The redshift limit of 0.366 is chosen to avoid any contamination due to redshift space distortion by assuming a maximum peculiar velocity of a galaxy in a massive cluster as
5000 km/s. To err on the side of caution and ensure no contamination from the fingers-of-gods effect, a conservative value of 10,000 km/s (corresponding to $\Delta z \sim 0.034$) has been chosen.
Figure~\ref{fig:mass-hmf} shows the mass and redshift distributions of the mock and WHL clusters.

\section{Methods: Modified Friends of Friends}
\label{sec:method}
To identify the superclusters and create a supercluster catalog, a modified friends-of-friends (mFoF) algorithm is used. The friends-of-friends (FoF) algorithm \citep{Huchra82,Martinez02} finds overdensities in 
the distribution of galaxies or clusters. In this algorithm, a linking length $l$ (threshold distance) is 
chosen. Two clusters are considered linked (part of the same supercluster) if the distance between 
them is less than or equal to $l$. This gives a system of clusters (a supercluster) with all the member clusters having 
distances to their natural neighbors $\le l$. We modify the FoF algorithm to account for the selection 
biases in the survey. The modified friends-of-friends algorithm makes use of Delaunay triangulation
for distance calculations and weighted linking length to find the member clusters. The procedure is 
described below in detail.

\subsection{Distances using Delaunay Triangulation}
A simple way is to calculate $O(N^2)$ distances for a distribution of $N$ clusters and then find $l$ for which the number of superclusters is maximum (see section~\ref{sec:linking length}). This method is computationally very expensive for large values of $N$. To increase the computational efficiency, Delaunay triangulation \citep{Delaunay34} is used to link a cluster to its nearest neighbors, and only the distances between the natural neighbors are considered. This reduces the calculation of $O(N^2)$ distances to $O(NlogN)$.

\newpage
\subsection{Radial Selection Weights}
\begin{figure*}
\centering
\includegraphics[width=\textwidth]{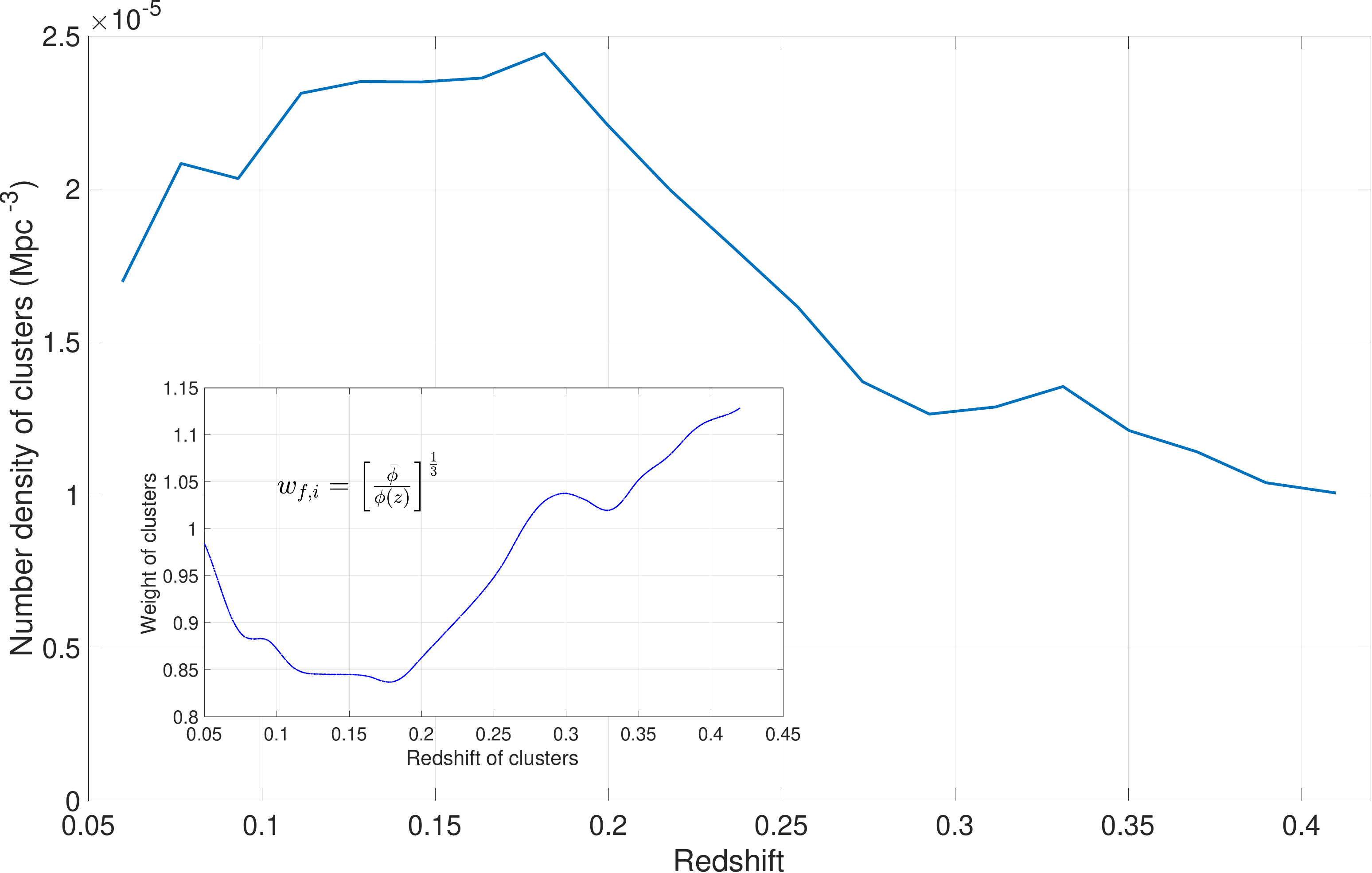}
\caption{Number density $\phi(z)$ of WHL clusters as a function of redshift within $0.05 \leq z \leq 0.42$. \textit{Inset}: Corresponding weights $w_{f,i}$ of the clusters as a function of redshift.}
\label{fig:WHLnofz}
\end{figure*}

As the number density of WHL clusters varies with redshift, the distribution of WHL clusters is not uniform in the radial direction. The number density $\phi(z)$ of WHL clusters is shown in Figure~\ref{fig:WHLnofz}.
The radial selection weight $w_{r,i}$ of the $i^{th}$ cluster at redshift $z$ is defined as
\begin{equation}
\centering
w_{r,i} = \frac{\bar{\phi}}{\phi(z)}
\label{eq:WHLradweight}
\end{equation}
where, $\bar{\phi}$ is the mean number density of the survey.

\subsection{Weighted Linking Length}
The number density of clusters varies with redshift (Figure~\ref{fig:WHLnofz}), necessitating the adjustment of the linking length according to this variation. Consequently, the linking lengths will be different for different pairs of clusters.
The radial selection weights are computed through three-dimensional density estimations. Therefore, the cube root of weights is to be taken to apply these weights on the one-dimensional linking length.
\begin{equation}
\centering
w_{f,i} = [w_{r,i}]^{\frac{1}{3}}
\label{eq:WHLrcweight}
\end{equation}
The final weights $w_{f,i}$ of all the clusters as a function of redshift are shown in Figure~\ref{fig:WHLnofz}.

Now, if $l_o$ is the selected linking length (as defined in Section~\ref{sec:linking length}), then the weighted linking length $l_{ij}$ between $i^{th}$ and $j^{th}$ clusters will be
\begin{equation}
\centering
l_{ij} = \frac{w_{f,i} + w_{f,j}}{2}~l_o
\label{eq:WHLlinklen}
\end{equation}

\subsection{Selecting the Linking Length}
\label{sec:linking length}
\begin{figure*}
\centering
\includegraphics[width=\textwidth]{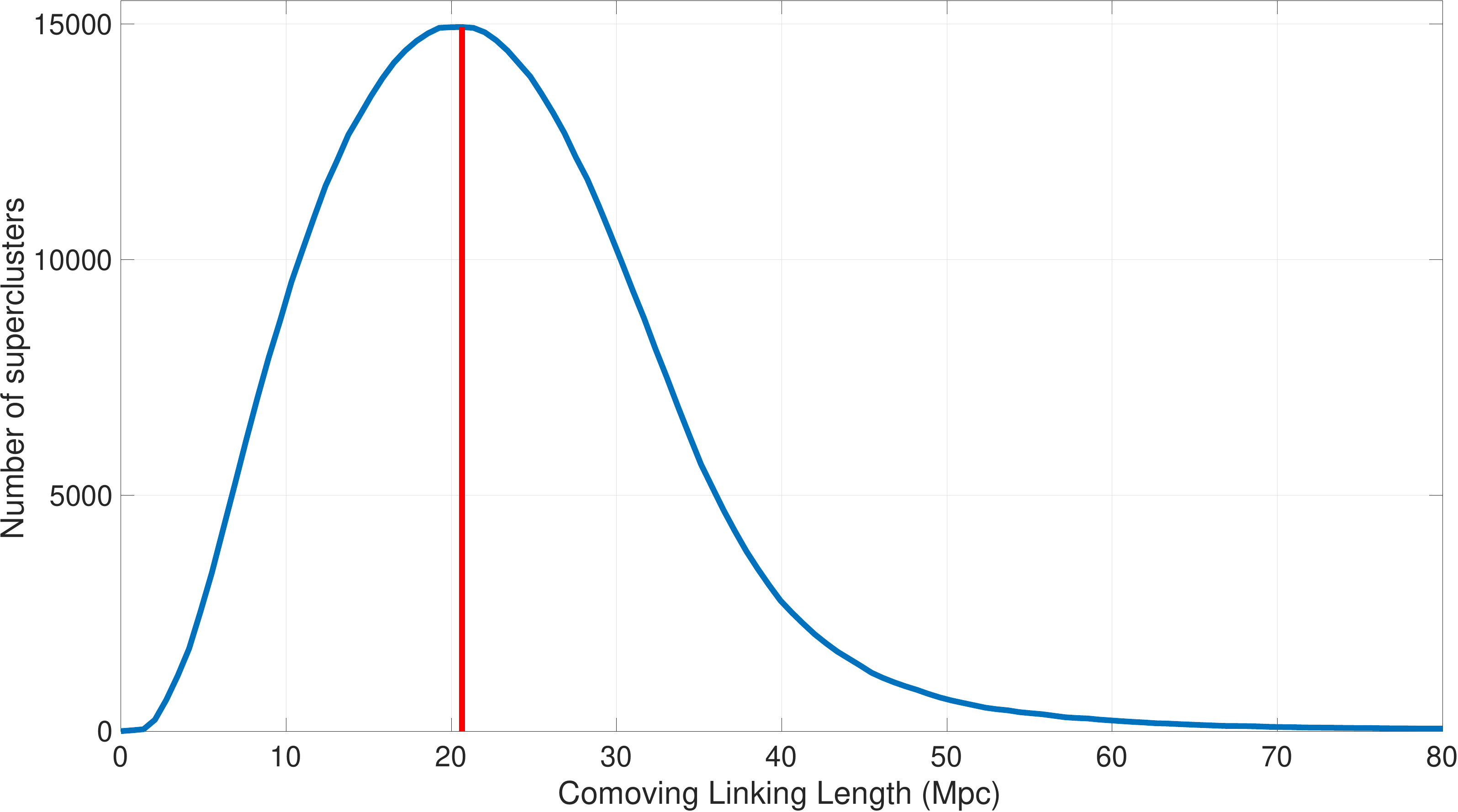}
\caption{Number of candidate superclusters as a function of unweighted comoving linking length $l_o$ (in blue). The red line shows the position $l_o = 20.65$ Mpc where the number of candidate superclusters is maximum.}
\label{fig:WHLll}
\end{figure*}

The $l_o$ is chosen to obtain the maximum number of superclusters for a given minimum number $N_{min}$ of clusters in a supercluster \citep{Chow-Martinez14,Bagchi17}. In literature, different FoF algorithms choose different $N_{min}$ for clustering analysis according to the data and the aim of the study. There are no fixed criteria for choosing $N_{min}$, and the choice changes the value of the linking length and, thus, the clustering properties.
In our analysis, we notice that selecting higher $N_{min}$ values and subsequently calculating the linking length leads to unstable results with the linking lengths varying significantly. We are limited by the number density of WHL clusters, and choosing a higher $N_{min}$ value increases the statistical noise. Moreover, higher $N_{min}$ gives a higher linking length value, which leads to the identification of some spurious superclusters exceeding 600 Mpc in size, comprised of up to 200 member clusters, and displaying negative density contrast values (indicating under-dense regions). This issue persists even when considering only the non-striped, contiguous survey region (discussed below). We mitigate this issue by choosing a lower value of $N_{min} = 2$, which
anchors the linking length estimation towards lower values, thereby preventing it from reaching percolation thresholds.
For our analysis, the chosen linking length ($l_o$) for the sample is 20.65 Mpc (as shown in Figure~\ref{fig:WHLll}), representing the maximum number of superclusters detected with $N_{min} = 2$.

The above method gives a range of masses of superclusters with at least two clusters in a supercluster, and hence, these can be regarded as only candidate superclusters.
The candidate superclusters with at least 10 member clusters are defined as superclusters in our catalog.
The majority ($\sim$ 80\%) of clusters in the WHL catalog have masses
$M_{200c}\gtrsim 10^{14} M_{\odot}$. With a minimum of 10 member clusters, the lower mass limit of a supercluster corresponds to $\sim 10^{15} M_{\odot}$.
Consequently, all the superclusters included in our catalog have masses $\gtrsim 10^{15} M_{\odot}$ which is consistent with the findings regarding supercluster masses presented in \citet{Araya-Melo09}.

To verify the robustness of our analysis results against the complex geometry of our sample's sky footprint, we conducted an additional check by examining only the contiguous regions of our sample footprint, excluding the striped regions.
We then performed the mFoF algorithm to detect superclusters. We found no significant deviations in the properties of the detected superclusters when comparing them to the findings presented in Section~\ref{sec:result} for the entire survey footprint (as illustrated in Figure~\ref{fig:whl-hr4-sel-fn}). This suggests that the inclusion of the striped regions in our analysis is justified. Furthermore,
the detected superclusters in striped regions will be useful for
individual studies. Only 10 of our catalog's 662 superclusters (Section~\ref{sec:result}) are found within the striped regions. We also note that the very massive Saraswati supercluster was found in the narrow Stripe 82 of the SDSS \citep{Bagchi17}.

\section{Properties of the Superclusters}
\label{sec:prop_super}
\subsection{Mass Estimation}
In order to calculate the total mass of a supercluster, we use the following method.
First, we calculate the bound mass of individual member clusters of a supercluster. Then, we add the bound masses of all the member clusters to get the total mass of the supercluster.

WH15 provides information about clusters' richness in their catalog, denoted as R$_{L\ast,500}$. This richness is the optical luminosity, $L_{500}$, measured in units of $L^*$, the evolved characteristic luminosity of galaxies in the r-band, within $R_{500c}$. WH15 also gives the correlation between the mass $M_{500c}$ (in the units of $10^{14}$ M$_{\odot}$) and richness R$_{L\ast,500}$ of the clusters as
\begin{equation}
\centering
\log M_{500c}=(1.08\pm0.02)\log R_{L\ast,500}-(1.37\pm0.02)
\label{eq:m500_rl}
\end{equation}
The mass uncertainty estimated by R$_{L\ast,500}$ is
$\sigma_{logM_{500c}} = 0.14$. To estimate the mass $M_{500c}$ of a cluster, we utilize R$_{L\ast,500}$ in conjunction with equation~\ref{eq:m500_rl}.

The relation between R$_{200c}$ and R$_{500c}$ is $R_{500c} \approx 0.65 \times R_{200c}$ for a given Navarro-Frenk-White (NFW) mass profile with a concentration parameter in the range 4 -- 8 \citep{Ettori09}.
And assuming a spherically symmetric distribution \textbf{($M_{{\Delta}c} = \Delta~\rho_c \frac{4\pi}{3} R_{{\Delta}c}^3$)} of halo density, M$_{200c}$ can be expressed as
\begin{equation}
\centering
M_{200c} = \frac{200}{500} \left(\frac{R_{200c}}{R_{500c}}\right)^{3} M_{500c}
\label{eq:m200_m500}
\end{equation}
Using equations~\ref{eq:m500_rl} and \ref{eq:m200_m500}, we get M$_{200c}$ for each cluster.
To account for the mass beyond the virial radius of a cluster, the virial mass (M$_{200c}$) of each member cluster is further scaled up by a factor of 2.2 to get its bound halo mass $M_{halo} \approx M_{5.6c}$.  In earlier studies, it is found that the  bound halo mass  of a cluster is  close to
 $M_{5.6c} \sim 2.2 \times M_{200c}$ \citep{Busha05,Rines06,Rines13}. 
 The sum of the bound halo masses of all member clusters gives the total mass of the supercluster.
Figure~\ref{fig:whl prop} shows the mass distribution of the superclusters.
\subsection{Size and Position}
\begin{figure*}
\centering
\gridline{\fig{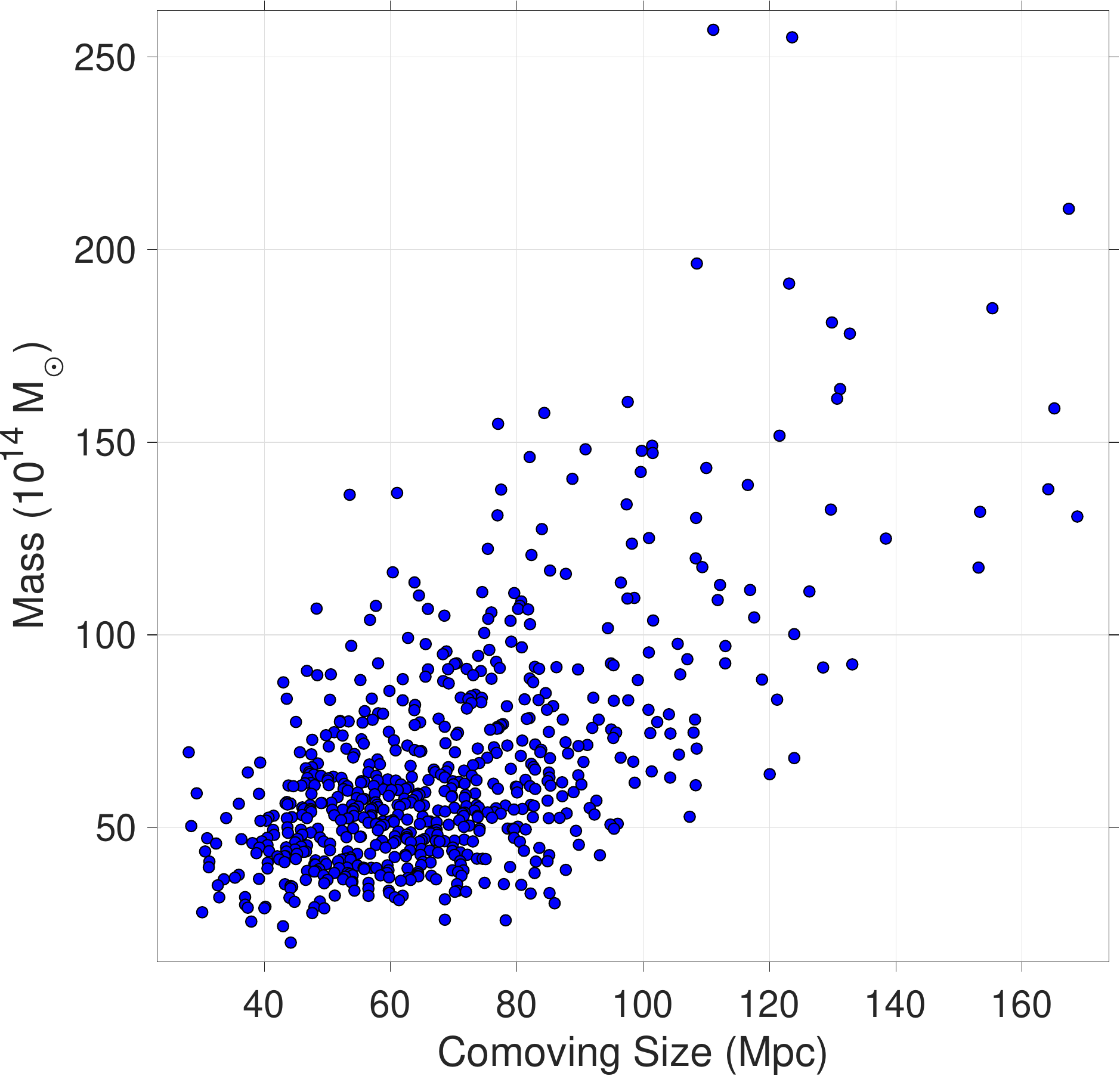}{0.32\textwidth}{(a)}
          \fig{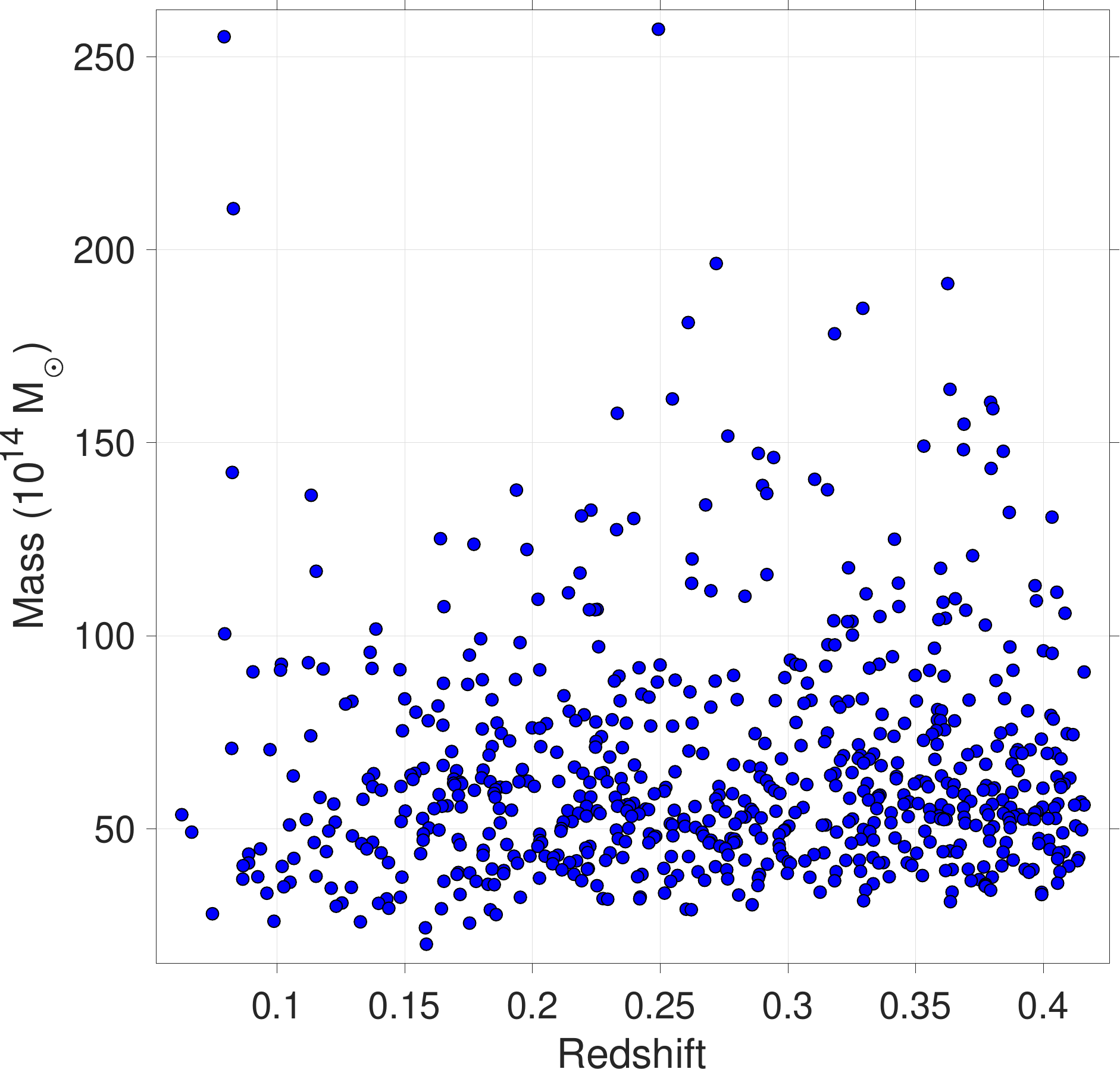}{0.325\textwidth}{(b)}
          \fig{delta_length_whl15_scatter.png}{0.3125\textwidth}{(c)}}
\caption{(a) Distribution of the superclusters' mass and size. (b) Distribution of the superclusters' mass and redshift. (c) Density contrast of the superclusters as a function of size. The blue dots represent the superclusters, the solid black line is a linear fit to the data points on the log-log plot, and the shaded black region denotes the associated errors in the fit. The slope of the fitted line is given at the top-right corner of the plot.}
\label{fig:whl prop}
\end{figure*}

The comoving linear size of a supercluster is calculated by measuring the maximum distance between pairs of member clusters of a supercluster. The comoving position $\mathbf{X}$ of the supercluster is taken as the virial mass-weighted average of the positions of member clusters,
\begin{equation}
\centering
\mathbf{X} = \frac{\sum_{i} M_{200c,i}~\mathbf{X}_i}{\sum_{i} M_{200c,i}}
\label{eq:WHLsupclustpos}
\end{equation}
where, $M_{200c,i}$ and $\mathbf{X}_i$ are the virial mass and comoving position of the $i^{th}$ member cluster. The distributions of the superclusters' size, mass, and redshift are shown in Figure~\ref{fig:whl prop}.

\newpage
\subsection{Density Contrast}
To estimate the average density contrast $\delta$ of a supercluster, we now need to estimate the volume
occupied by a supercluster. Since these superclusters do not have any specific shape, we estimate the volume by fitting a convex hull to its member clusters. A convex hull of a set of points \textbf{P} in 3D Euclidean space is the convex surface (envelope) of the minimum possible volume on \textbf{P}.
The volume of a supercluster is much larger than the volume of a single cluster. Hence, treating clusters as points or spherical objects with fixed radii has negligible impact on the estimated supercluster volumes.
We use the Qhull algorithm \citep{Barber96} to construct the convex hull and to calculate its volume. Once we have the mass and volume, we can calculate the density of the supercluster. The corresponding matter density contrast, therefore, will be given by,
\begin{equation}
 \delta = \frac{\rho_{SC}}{{\rho_{m}}} - 1
\label{dens_contrast}
\end{equation}
where $\rho_{SC}$ is the mass density of the supercluster and $\rho_{m}$ is the background matter density at the redshift ($z$) of the supercluster,
\begin{equation}
 \rho_{m} = \frac{3 H_0^2 \Omega_{m} (1+z)^3}{8 \pi G}
\label{dens_back}
\end{equation}
Figure~\ref{fig:whl prop} shows the density contrast of the superclusters.

\section{Results}
\label{sec:result_all}

\subsection{Observation}
\label{sec:result}
\subsubsection{WHL Superclusters}
\begin{figure*}
\centering
\includegraphics[width=\textwidth]{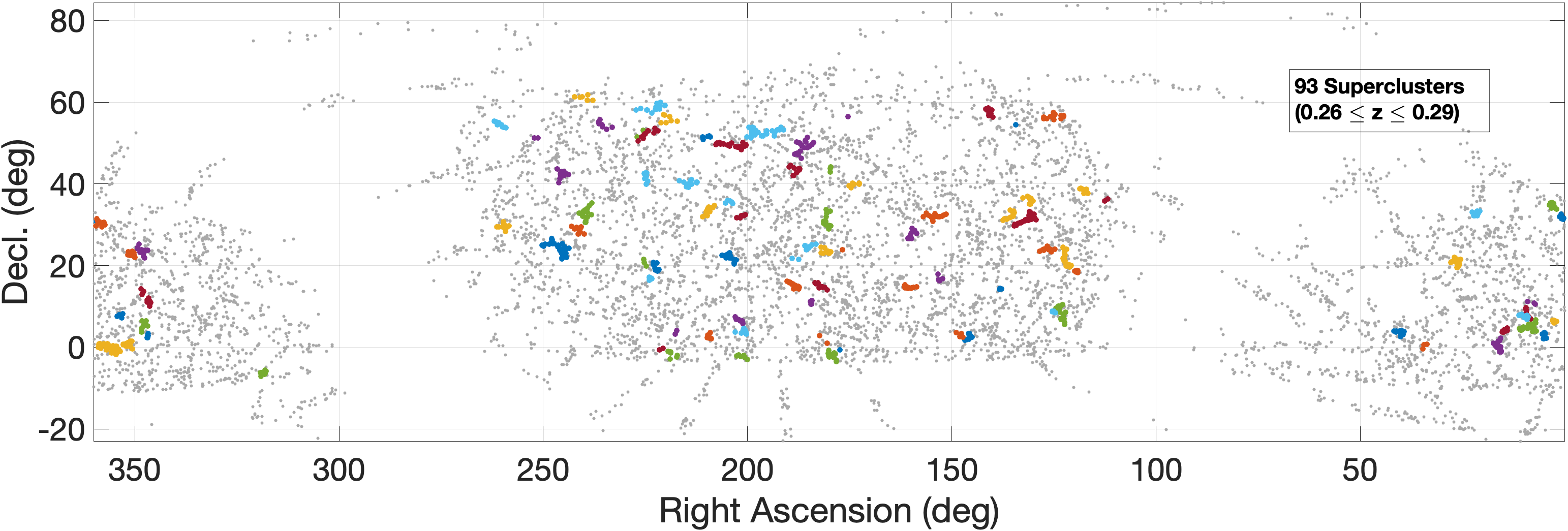}
\caption{Sky plane distribution of the superclusters within $0.26 \leq z \leq 0.29$. Dots represent the WHL clusters within the redshift slice. Grey dots are not a part of any supercluster, and the other colored dots represent the member clusters of 93 superclusters within the redshift slice.}
\label{fig:redshift_slice}
\end{figure*}

The mFoF algorithm with at least 10 member clusters gives a total of 662 superclusters in the redshift range of $0.05 \leq z \leq 0.42$. To the best of our knowledge, this is the
largest supercluster catalog constructed from the data of clusters in this redshift range. Around $12\%$ clusters (9895 clusters) from the WHL catalog considered for our work reside in a supercluster environment, and around $28\%$ (183 superclusters) of the superclusters have at least one Abell cluster \citep{Abell89}
in them. The presence of Abell clusters in many superclusters (for example, Shapley supercluster \citep{Shapley1930,Raychaudhury89}, Sloan Great Wall, Saraswati supercluster, and Corona Borealis supercluster)
demonstrates the robustness of our supercluster catalog.
Figure~\ref{fig:redshift_slice} shows the distribution of the superclusters in the redshift slice $0.26 \leq z \leq 0.29$.
All the WHL clusters within the redshift slice are represented by dots. Grey dots represent the clusters that are not part of any supercluster. The other colored dots represent member clusters of 93 superclusters within the redshift slice. The supercluster in yellow at $350^{\circ} \leq RA \leq 359^{\circ}$ and $Decl. \sim 0^{\circ}$ is the Saraswati supercluster.

The sky coordinates, redshifts, and properties
of the superclusters, arranged in the decreasing order of their masses, in our catalog are listed in Table~\ref{tab:WHLsupclustcat}.
The distributions of superclusters' mass, size, and density contrast are shown in Figures~\ref{fig:whl prop}.
\setlength{\tabcolsep}{8pt}
\begin{table*}
\centering
\caption{Properties of the sample of 662 superclusters in the decreasing order of their masses. BCG stands for brightest cluster galaxy. Columns represent: (1) Supercluster identification number in our catalog, (2) Right ascension, (3) Declination, (4) Right ascension of the most massive member cluster's BCG, (5) Declination of the most massive member cluster's BCG, (6) Redshift, (7) Linear comoving size, (8) Total Mass, (9) Number of member clusters, and (10) Number of Abell clusters present.\label{tab:WHLsupclustcat}}
\begin{tabular}{|c|c|c|c|c|c|c|c|c|c|}
\hline
\hline
\textbf{SCl} & \textbf{RA} & \textbf{Decl.} & \textbf{RA\_BCG} & \textbf{Decl\_BCG} & \textbf{z} & \textbf{Size} & \textbf{Mass} & \textbf{N$_{mem}$} & \textbf{N$_{Abells}$} \\
 & \textbf{(deg)} & \textbf{(deg)} & \textbf{(deg)} & \textbf{(deg)} &  & \textbf{(Mpc)} & \textbf{($10^{14}~M_{\odot}$)} &  &  \\
(1) & (2) & (3) & (4) & (5) & (6) & (7) & (8) & (9) & (10) \\
\hline
1 & 209.6316 & 3.3618 & 210.2586 & 2.8785 & 0.2493 & 111.10 & 257.106 & 54 & 2 \\
2 & 235.1322 & 28.2855 & 239.5833 & 27.2334 & 0.0792 & 123.59 & 255.164 & 57 & 7 \\
3 & 200.3919 & 0.2291 & 202.7959 & -1.7273 & 0.0828 & 167.40 & 210.629 & 41 & 10 \\
4 & 356.0263 & -0.0101 & 355.9481 & 0.2567 & 0.2719 & 108.50 & 196.421 & 38 & 1 \\
5 & 3.2246 & 8.3751 & 3.1893 & 8.3964 & 0.3625 & 123.10 & 191.222 & 44 & 0 \\
\hline
\end{tabular}
\tablecomments{This table will be available in its entirety in a machine-readable format in the online version of the journal. A portion is shown here for guidance regarding its form and content. The full table will also be available on CDS.}
\end{table*}

Apart from a few superclusters, which are the rediscovery of the previously known superclusters, most of the 
superclusters reported here are newly discovered.
For e.g., Saraswati supercluster (SCl 4 and SCl 231) and the Sloan 
Great Wall (SCl 3 and SCl 22) are identified by our algorithm (see section~\ref{sec:mast_mass_sc}).

The median size and mass of the superclusters in our catalog are $\sim$ 65 Mpc and $\sim 6 \times 10^{15}~M_{\odot}$, respectively.
It is to be noted that the superclusters listed in our catalog might be even larger and more massive than currently indicated. We estimate the masses of these superclusters by adding the bound masses of their member clusters. However, matter also exists between these clusters within a supercluster. Currently, we do not have an estimation of the mass attributed to this inter-cluster matter. Consequently, the actual mass of the superclusters may be higher than what is indicated by our calculations. In this context, \citet{Bagchi17}
found that if one assumes that the average  density of diffuse matter (baryonic plus dark matter) dispersed between component (member) clusters is minimally at the cosmological matter density, then there is an increase in the total mass of Saraswati supercluster at least by a factor of two.
Furthermore, it is worth considering that the superclusters situated at the edges of the SDSS survey footprint and those near the redshift limits of our sample might only be partially identified in terms of their true extent. Therefore, factors such as survey boundaries and limitations imposed by redshift constraints could potentially impact our ability to capture the entirety of these superclusters fully.
A very deep and dedicated spectroscopic survey is needed to unravel their true extent and content. This sample will also be useful for 
comparing similar studies using numerical simulations. Also, since the catalog has a relatively wide 
range of redshift coverage spanning $\sim$ 4 Gyr, the study of the evolution of the properties of galaxies in the supercluster 
environments can be explored.

Our supercluster catalog differs in some ways from two other catalogs: the one by \citet{Chow-Martinez14} and the other by \citet{Liivamagi12}.
In the catalog by \citet{Chow-Martinez14}, the redshift coverage is less, up to $z \leq 0.15$. However, their sky coverage is significantly larger than ours, encompassing almost the entire sky except for the Galactic disk region. To identify the superclusters, they implemented a tunable friends-of-friends algorithm based on the selection functions, applied to the Abell/ACO clusters \citep{Abell58, Abell89}. The number of member clusters in their catalog varies from 2 to 42, and the sizes of the identified superclusters range from $<$ 1 Mpc to $\sim$ 185 Mpc.
On the other hand, the supercluster catalog of \citet{Liivamagi12} has a redshift coverage of $0.02 \leq z \leq 0.5$, but its sky coverage is less than 50\% compared to our supercluster catalog. They took a different approach, identifying supercluster regions using the luminosity density field generated from the SDSS's DR7 \citep{Abazajian2009} spectroscopic galaxy data. Applying an adaptive density threshold on this field, they found supercluster sizes ranging from $\sim$ 22 Mpc to $\sim$ 260 Mpc.
In contrast to these catalogs, our supercluster catalog provides additional information on mass estimates for the identified superclusters. Along with a wide sky coverage and a comparatively large redshift range of our catalog, this additional information makes it a useful resource for further research and understanding of superclusters.

We also find a correlation between the density contrast and the size of the superclusters. It is close to a power law with an index, $\alpha \sim -2$. The correlation (Figure~\ref{fig:whl prop}) is,
\begin{equation}
\centering
\log \delta = (-2.14 \pm 0.07)~\log L + (4.95 \pm 0.13)
\end{equation}
where, $\delta$ is the density contrast and $L$ is the size of the supercluster.
In our catalog, we notice that this correlation arises from the dependency of mass ($\mathrm{M}$) and volume ($\mathrm{V}$) of the superclusters on its size ($\mathrm{M} \propto L^{0.73}$ and $\mathrm{V} \propto L^{2.64}$).
It may arise due to the morphology distribution of the superclusters or due to the evolution of the superclusters/structures with time \citep{Teerikorpi15}. We leave this aspect to explore in future work.

\subsubsection{Notes on the Five Most Massive Superclusters}
\label{sec:mast_mass_sc}
Here, we highlight the five most massive superclusters in our supercluster catalog.
Figure~\ref{fig:scl1-5} shows the sky distribution of these five superclusters. Figure~\ref{fig:optical_images} shows the optical images of
the most massive member clusters of the top five superclusters in Table~\ref{tab:WHLsupclustcat} from DESI Legacy Imaging Surveys (DR9)\footnote{\url{https://www.legacysurvey.org}}. These are one of the richest clusters in the WHL cluster catalog. The WHL cluster catalog places the cluster ZwCl 2341.1+0000 (Figure~\ref{fig:optical_images} - SCl 4) as the most massive cluster of SCl 4 (Saraswati) supercluster. However, Abell 2631 (also shown in Figure~\ref{fig:optical_images}) is the most massive cluster of Saraswati \citep{Bagchi17,Monteiro-Oliveira21}. Zwcl 2341.1+0000 is a multiple merger system \citep{Bagchi02,van_Weeren09,Boschin13,Viral22}, which is why the mass estimation of this cluster is slightly higher. The individual notes on the top five superclusters are given below :

\textbf{SCl 1 - new discovery} :
The most massive supercluster (SCl 1) is found at redshift $\sim 0.25$ with mass $\sim 2.57 \times 10^{16}~M_{\odot}$ and size $\sim 111$ Mpc. This supercluster contains 54 member clusters, including Abell 1835 and Abell 1801. Abell 1835 is the most massive member cluster of SCl 1 (see Figure~\ref{fig:optical_images}). Abell 1835 is a massive cool core cluster \citep{Ueda20117} with a very high mass of $\sim$\,10$^{15}~\mathrm{M}_{\sun}$
\citep{Schmidt2001,LaRoque2006} and a radio mini-halo \citep{Kale2015}. SCl 1 is a newly discovered supercluster and is the most massive at this redshift.
We propose to name SCl 1 as \textit{Einasto Supercluster} in honor of Prof.\ Jaan Einasto\footnote{Prof.\ Jaan Einasto is one of the pioneers of the cosmic web. He is one of the first to start systematic studies of superclusters and still actively contributes to the field.}.

\begin{figure*}
\centering
\includegraphics[width=\textwidth]{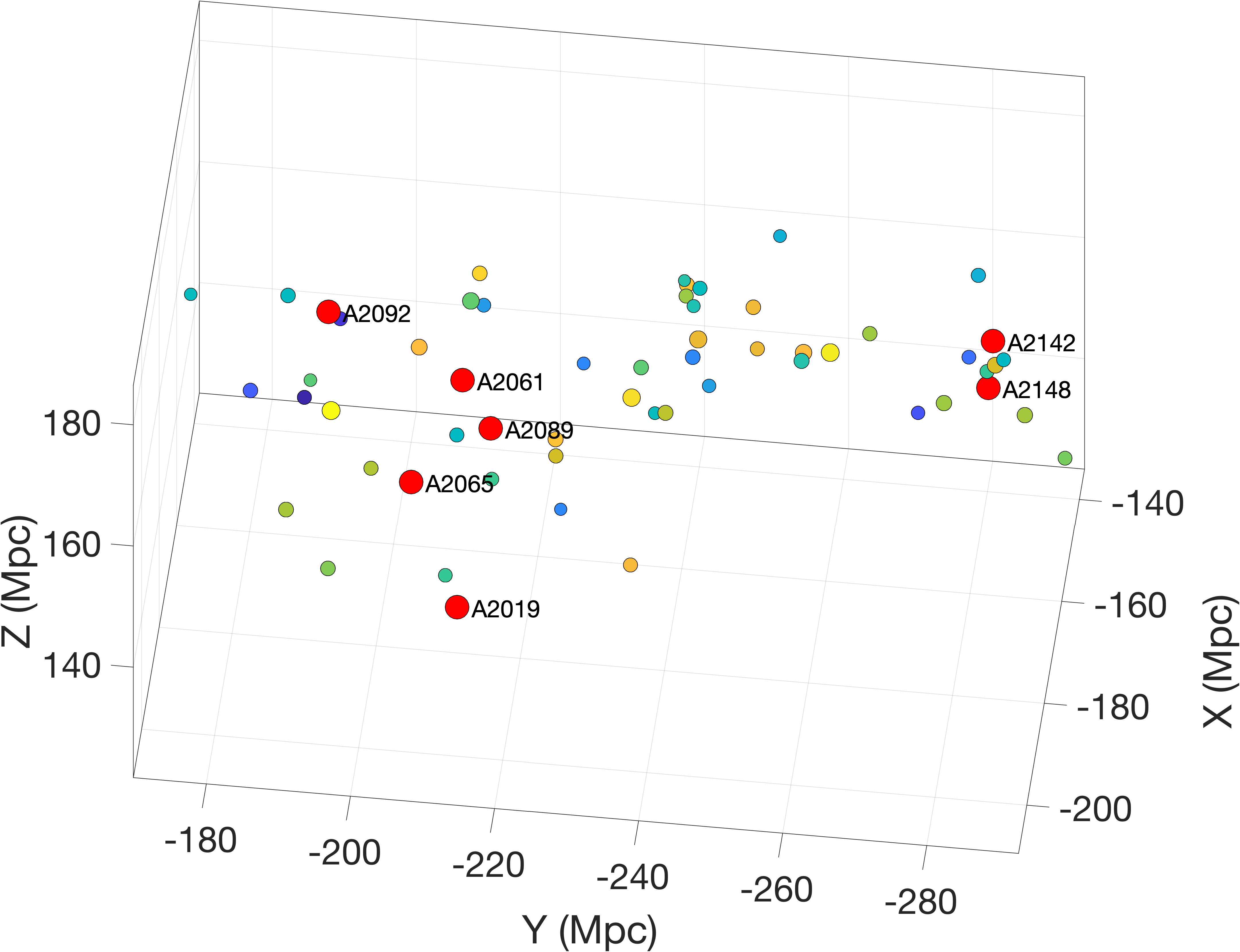}
\caption{The three-dimensional distribution of clusters in the richest supercluster of our catalog (SCl 2), the Corona Borealis supercluster. Colors represent the mass of the clusters, with blue representing the less massive cluster and yellow the most massive. Red spheres represent the Abell clusters; their Abell IDs are written next to them. There are a total of 57 clusters in SCl 2.}
\label{fig:WHLmostmassive}
\end{figure*}

\textbf{Corona Borealis Supercluster - rediscovery} :
In our catalog, the richest (maximum number of member clusters) supercluster is SCl 2 with 57 member clusters, including 7 Abell clusters,
at redshift $z \approx 0.08$, mass $M \approx  2.55\times 10^{16}\mathrm{M}_{\sun}$,
and size $\approx 124$ Mpc. It contains most of the Corona Borealis (CB) supercluster \citep{Einasto21}.
Figure~\ref{fig:WHLmostmassive} shows the three-dimensional distribution of 57 clusters of SCl 2 in the comoving coordinate space.
\citet{Einasto21} showed that the CB supercluster consists of two parts, weakly connected by a chain of galaxies and poor groups. They found that these parts will separate during future evolution in physical space and form two separate structures. One of them, which contains Abell clusters A2065, A2061/A2067, and A2089, will be one of the most massive bound systems in the local Universe \citep{Einasto21}.
Our algorithm combines the CB supercluster with the
supercluster A2142
(see \citealt{EinastoM15,EinastoM18,EinastoM20} for details and references).
As a result, the total mass of this double supercluster system in our
catalog is twice as high
as the sum of masses of the CB and A2142 superclusters according to the
estimates in \citet{EinastoM15,EinastoM21}. Also, A2142 is the most massive cluster in the CB supercluster in our catalog.
It is worth mentioning that the inner structure of the CB is complicated. It consists of multiple clusters connected by low-density filaments, resembling a group of spiders  \citep[called a multispider in a  morphological classification
of superclusters in][]{EinastoM11_DR7SCl}. Overall, the CB has a highly elongated shape resembling a horse-shoe \citep{EinastoM11_DR7SCl}.

\textbf{Sloan Great Wall - rediscovery} :
The richest supercluster complex in the local Universe, the Sloan Great Wall (SGW), is represented by two superclusters in our catalog, SCl 3 and SCl 22 (Figure~\ref{fig:scl1-5}).
In \citet{Park12}, the spectroscopic galaxy data from the SDSS is utilized to identify nearby structures. Due to this data's higher density than our sparser cluster data, these superclusters have been identified as a single system \citep{Park12}.
The morphology of SCl 3, the richest SGW supercluster,
resembles a multi-branching filament  with rich galaxy clusters connected by
a small number of filaments.
The second richest supercluster in the SGW, SCl 22 in our catalog, is
multispider, in which clusters and groups are connected by a large number of filaments. During
future evolution in physical space,
both these superclusters  will fall apart and form smaller superclusters
\citep{EinastoM16}.
SGW superclusters differ in galaxy content and morphology, suggesting
that they have had
different evolution \citep{EinastoM10,EinastoM11_SGW}.

\textbf{Saraswati Supercluster - rediscovery} :
Saraswati supercluster \citep{Bagchi17} is a massive supercluster surrounded by large voids at a redshift $\sim 0.28$. In our catalog, it is identified with SCl 4 and SCl 231 (Figure~\ref{fig:scl1-5}). The total number of member clusters is 51 (38 clusters in SCl 4 and 13 in SCl 231).
In \citet{Bagchi17}, a high-density spectroscopic galaxy sample was used to identify the Saraswati supercluster. This sample allowed them to observe a continuous structure encompassing SCl 4 and SCl 231. The combined mass of SCl 4 and SCl 231 is $\sim2.6 \times 10^{16}$ M$_{\odot}$, slightly higher but still within the same order of magnitude as the value reported in \citet{Bagchi17}. The increased mass is attributed to the inclusion of more member clusters, made possible by the wider sky-area coverage of the WHL cluster catalog compared to the spectroscopic galaxy sample used in \citet{Bagchi17}.
The most massive cluster of Saraswati is Abell 2631, and the second most massive cluster is ZwCl 2341+0000. \citet{Bagchi17} showed that the central core region within a radius $\sim 20$ Mpc of this supercluster, including Abell 2631, is gravitationally bound. The total size of this supercluster is $\sim 200$ Mpc, which is the largest supercluster at this redshift.

\textbf{SCl 5 - new discovery} :
SCl 5 is our catalog's fifth most massive supercluster, containing 44 clusters at a redshift $\sim 0.36$. Its mass and size are $\sim 1.9 \times 10^{16}$ M$_{\odot}$ and $\sim 123$ Mpc, respectively. It does not contain any Abell cluster. It is also a newly discovered supercluster.
 

\subsubsection{Effect of supercluster environment on the clusters}
\begin{figure*}
\centering
\gridline{\fig{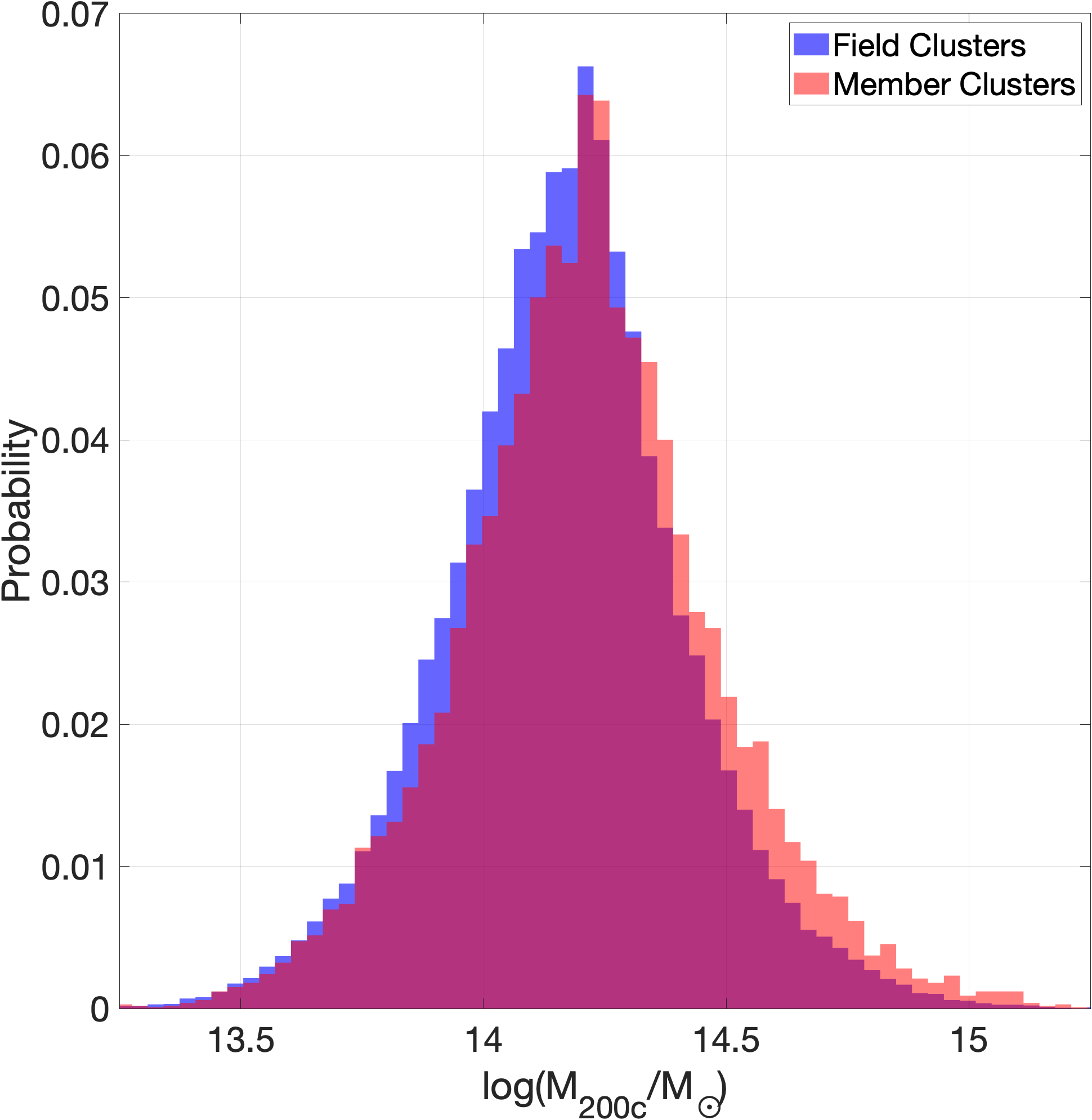}{0.475\textwidth}{(a)}
          \fig{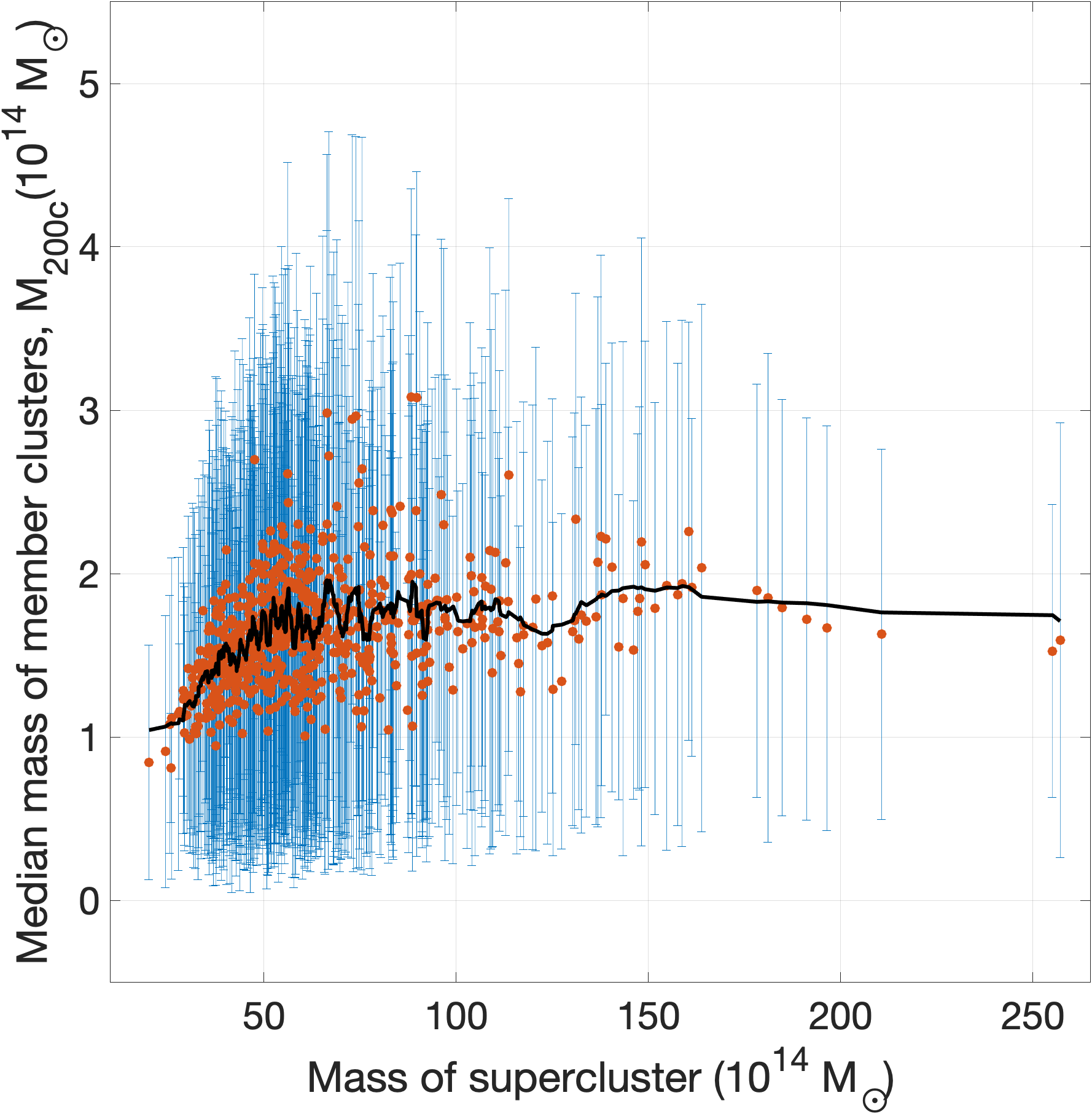}{0.485\textwidth}{(b)}}
\caption{
(a) Normalized histograms of the masses of clusters within superclusters (red histogram) and in the field (blue histogram). (b) Median masses (red dots) of the member clusters of supercluster as a function of their total mass. Blue error bars denote the quartile values of the member cluster masses above and below the median values, and the solid black line is the running mean of red dots.
}
\label{fig:clust_mass_bias}
\end{figure*}

The cosmic environment affects the growth and evolution of galaxies and clusters. To explore whether the supercluster environment plays any role in the evolution of clusters, we study the mass distribution of clusters in superclusters and in the field (clusters that are not members of any supercluster). The left panel of Figure~\ref{fig:clust_mass_bias} shows the normalized mass distributions of member clusters of superclusters and the field clusters. It shows a slight mass bias of finding massive clusters in a supercluster environment than in the field. Similarly, there is a slight deficiency of less massive clusters within superclusters. In other words, the probability of randomly picking a massive cluster is slightly higher within a supercluster environment than in the field. And the probability of randomly picking a less massive cluster is slightly higher in the field than in a supercluster region. To see the difference between the two distributions, we perform a two-sample Kolmogorov-Smirnov (KS) test \citep{Kolmogorov33,Smirnov48}. The null hypothesis that the two samples are drawn from the same population is rejected with a p-value of $3.7 \times 10^{-52}$.
This shows that the supercluster environment affects the evolution of clusters.

We also explore the mass distribution of the member clusters within each supercluster. The right panel of Figure~\ref{fig:clust_mass_bias} shows the median values (red dots) of the mass of clusters in a supercluster as a function of the total mass of the supercluster. Here, despite a large scatter, we see a trend indicating that the low-mass superclusters (masses $\lesssim 5 \times 10^{15}$ M$_{\odot}$) host low-mass member clusters. This may suggest that the growth of clusters is higher in high-mass superclusters. However, this growth in mass is not very high, the median value goes from $\sim 1 \times 10^{14}$ M$_{\odot}$ to $\sim 1.8 \times 10^{14}$ M$_{\odot}$.

\subsection{Simulation}
\label{sec:result_sim}
\subsubsection{Comparison with Simulation}
\begin{figure*}
\centering
\includegraphics[trim={5.4cm 1.9cm 5.4cm 2cm},clip,width=\textwidth]{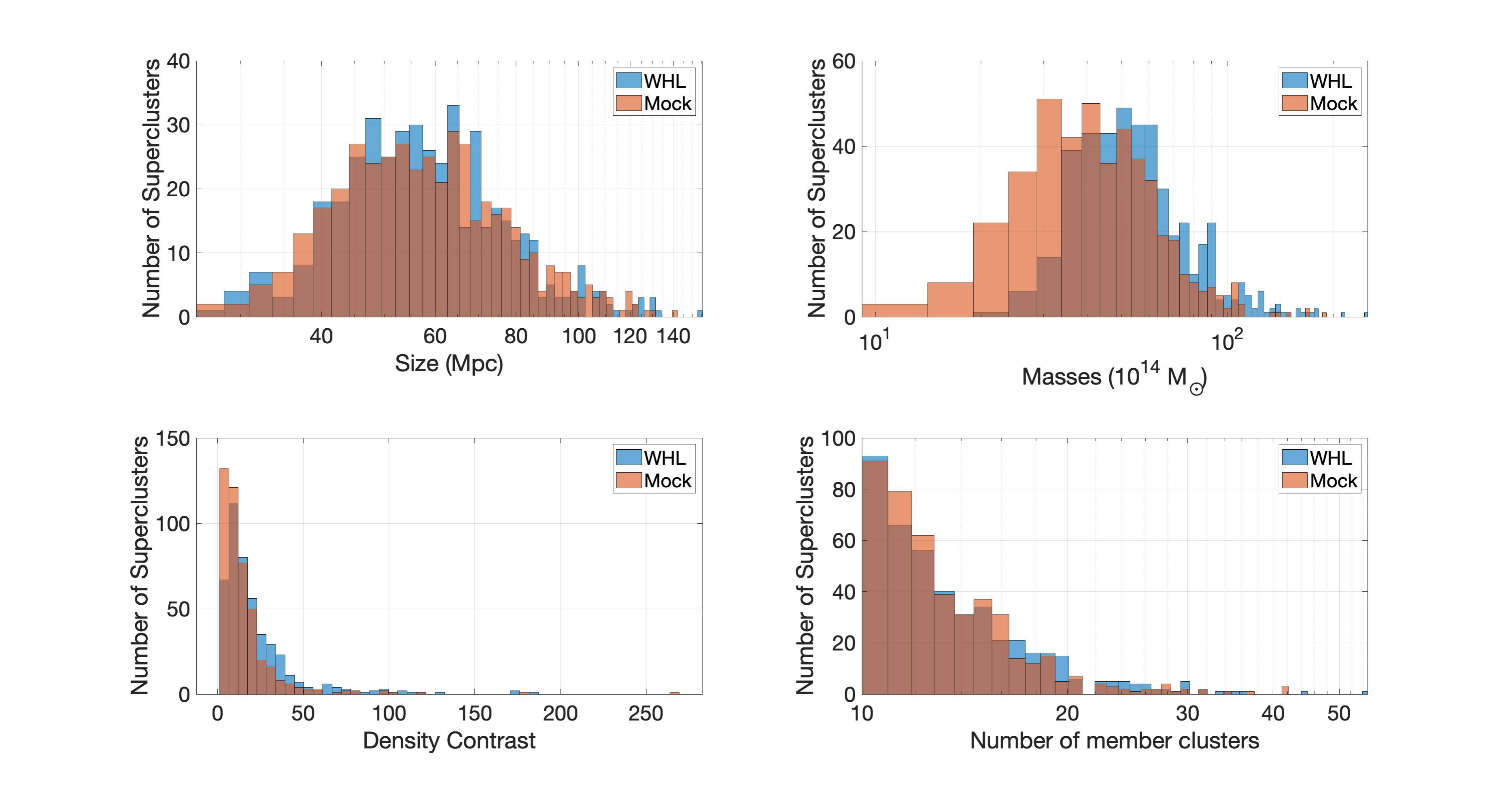}
\caption{Distributions of the size, mass, density contrast, and the number of member clusters of the superclusters in WHL and HR4 within $0.05 \le z \le 0.366$.}
\label{fig:whl_hr4_prop}
\end{figure*}

\begin{figure*}
\centering
\gridline{\fig{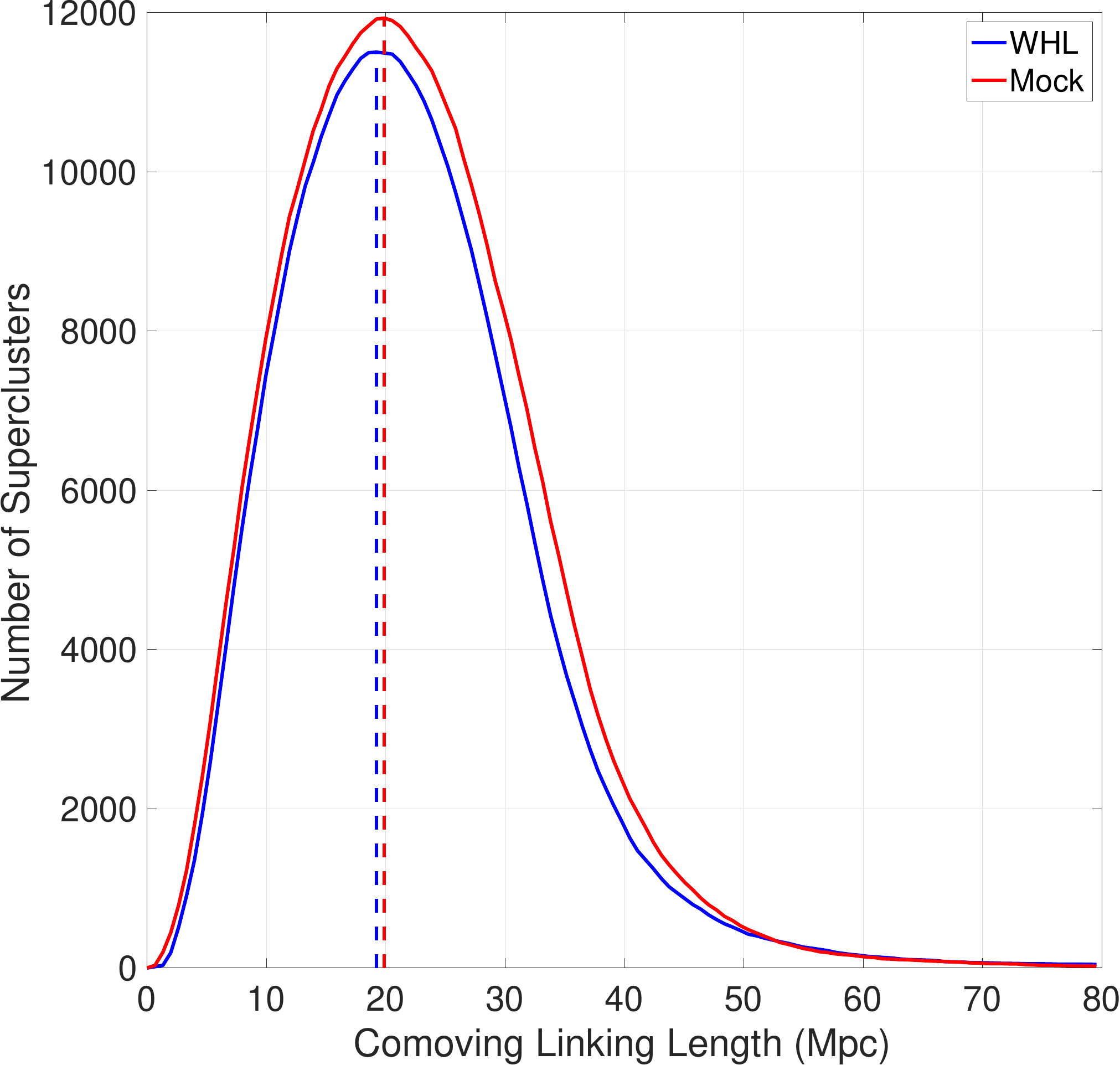}{0.5\textwidth}{(a)}
          \fig{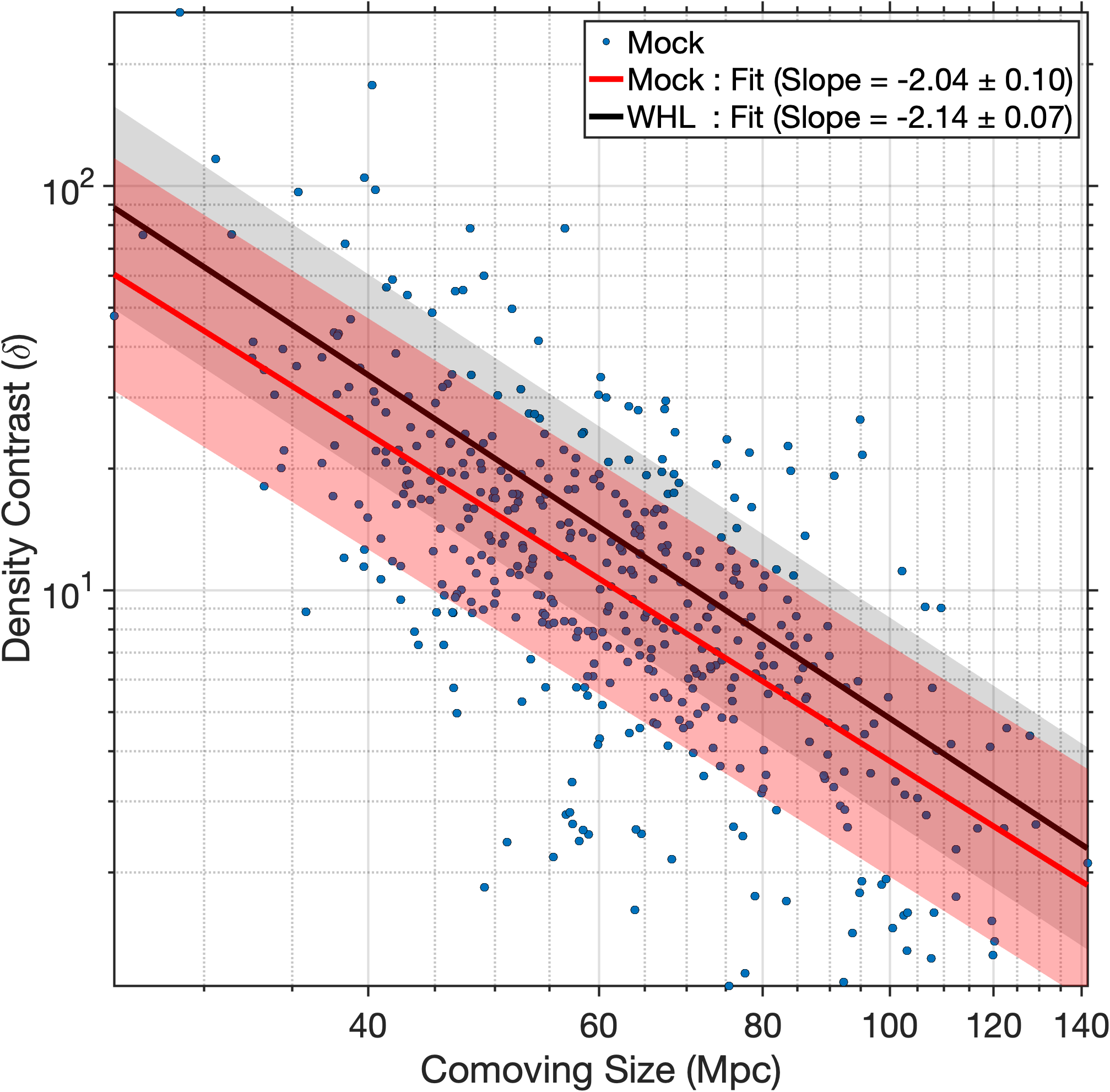}{0.485\textwidth}{(b)}}
\caption{(a) Number of candidate superclusters for different linking lengths $l_0$ for WHL and mocks. The curves peak at 19.22 Mpc and 19.88 Mpc for WHL and mocks, respectively, and are represented by dotted lines.  (b) Density contrast of the mock superclusters as a function of
size. Blue dots represent the mock superclusters, the red line is a linear fit to the data points on the log-log plot, and the shaded red region denotes
the associated errors in the fit. The slope of the fitted line is given at the top-right corner of the plot. The black line and the shaded black region denote the corresponding fit and the associated errors of WHL superclusters presented in Figure~\ref{fig:whl prop}(c).}
\label{fig:whl_hr4_ll_delta_size}
\end{figure*}

We compare our results with the mock superclusters extracted from the mock WHL clusters of the HR4 simulation (as described in section~\ref{sec:mock}). To compare the redshift coverage of mock clusters, we restrict our observational data of clusters to $0.05 \le z \le 0.366$. We then apply the same mFoF algorithm on the observed and mock clusters to identify superclusters. The results are summarized in Figure~\ref{fig:whl_hr4_prop}, Figure~\ref{fig:whl_hr4_ll_delta_size} and Table~\ref{tab:whl_hr4}.

The distributions of the size, density contrast, and the number of member clusters match fairly well, but there is a slight difference in the distribution of masses of superclusters (Figure~\ref{fig:whl_hr4_prop}). The difference in the mass distributions is because of the deficiency of massive clusters in the mock WHL clusters made from the HR4 simulation (see section~\ref{sec:mock}).

The linking lengths for WHL and mock clusters are 19.22 Mpc and 19.88 Mpc, respectively (Figure~\ref{fig:whl_hr4_ll_delta_size}). This difference of 0.66 Mpc is acceptable as it is much less than the mean cluster separation ($\sim 40$ Mpc) and less than the mean diameter of a cluster ($\sim 2$ Mpc).

We also see the correlation between the density contrast and the size of a supercluster in the mock superclusters (Figure~\ref{fig:whl_hr4_ll_delta_size}). The value of the slope here is $\sim -2.04 \pm 0.10$.

\setlength{\tabcolsep}{5pt}
\begin{table*}
\centering
\caption{Summary of the properties of the observed and mock superclusters.\label{tab:whl_hr4}}
\begin{tabular}{|c|c|c|c|}
\hline
\hline
 & \textbf{WHL} & \textbf{WHL} & \textbf{Mock} \\
 & \textbf{(0.05 $\le$ z $\le$ 0.42)} & \textbf{(0.05 $\le$ z $\le$ 0.366)} & \textbf{(0.05 $\le$ z $\le$ 0.366)} \\
\hline
\hline
Linking length $l_o$ (Mpc) & 20.65 & 19.22 & 19.88\\
Number of Superclusters & 662 & 456 & 451\\
Median Mass ($10^{14}$ M$_{\odot}$) & 57.80 & 55.97 & 46.43\\
Median Size (Mpc) & 64.87 & 60.49 & 60.83\\
Median Density Contrast & 11.64 & 15.40 & 10.88\\
Median No. of Members & 13 & 13 & 12\\
\hline
\end{tabular}
\end{table*}

\subsubsection{Peculiar Velocities of Members in the Mock Superclusters}
\label{sec:pec_vel}
\begin{figure*}
    \centering
    \includegraphics[width=\textwidth]{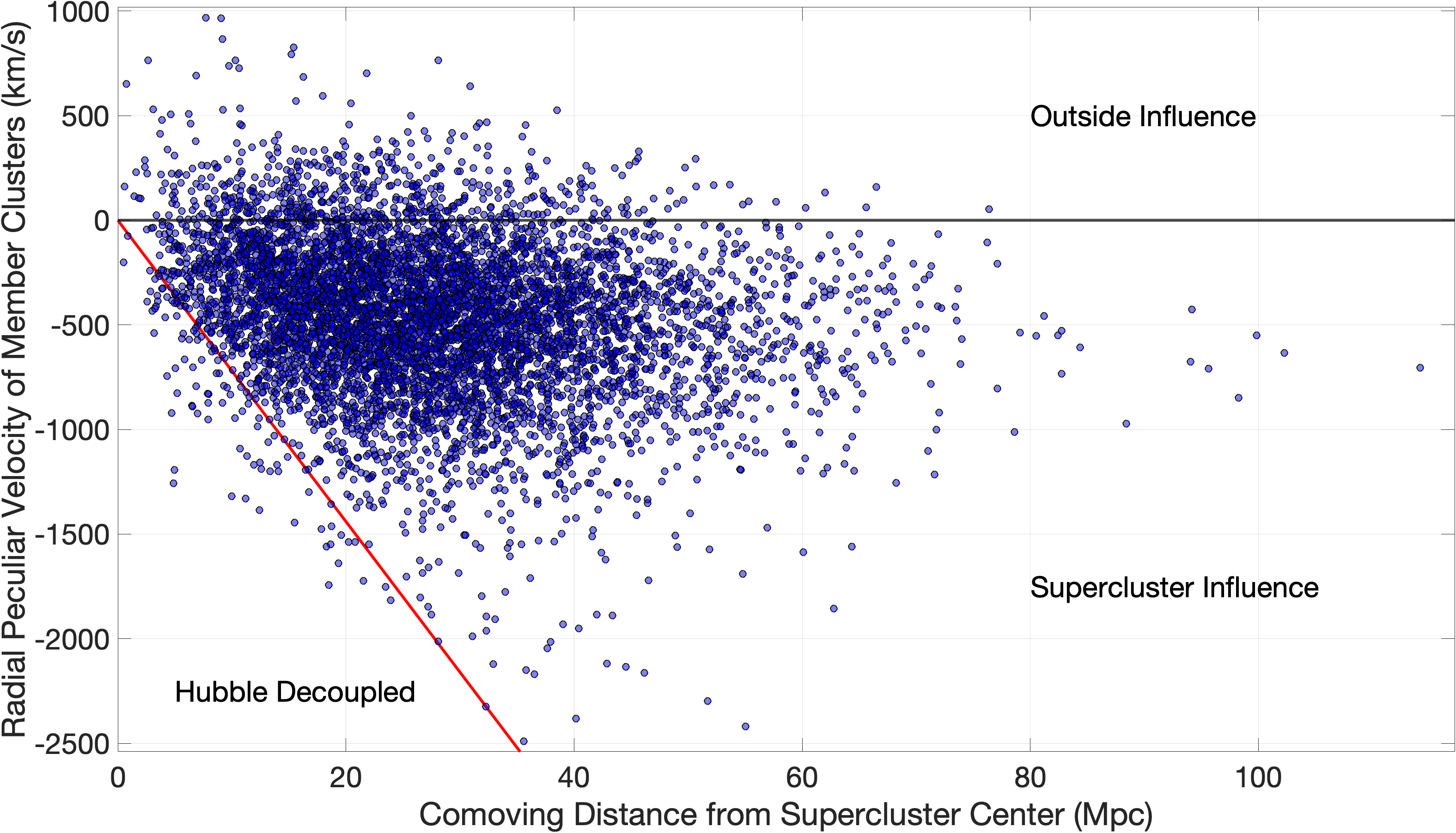}
    \caption{Radial peculiar velocities of member clusters of all (stacked) superclusters as a function of their comoving distance from the supercluster center. The black line represents a peculiar velocity of 0 km/s, and the red line represents all the peculiar velocities equal and opposite to the Hubble flow as seen from the center. The three regions, divided by red and black lines, are \textit{Hubble Decoupled}: Members that have decoupled from the Hubble flow, \textit{Supercluster Influence}: Members that have not decoupled from Hubble flow but have slowed down in their Hubble expansion and \textit{Outside Influence}: Member clusters that have their velocities greater than the Hubble flow and therefore have gravitational influence from regions outside of the supercluster.}
    \label{fig:pec_vel}
\end{figure*}

\begin{figure*}
    \centering
    \includegraphics[width=\textwidth]{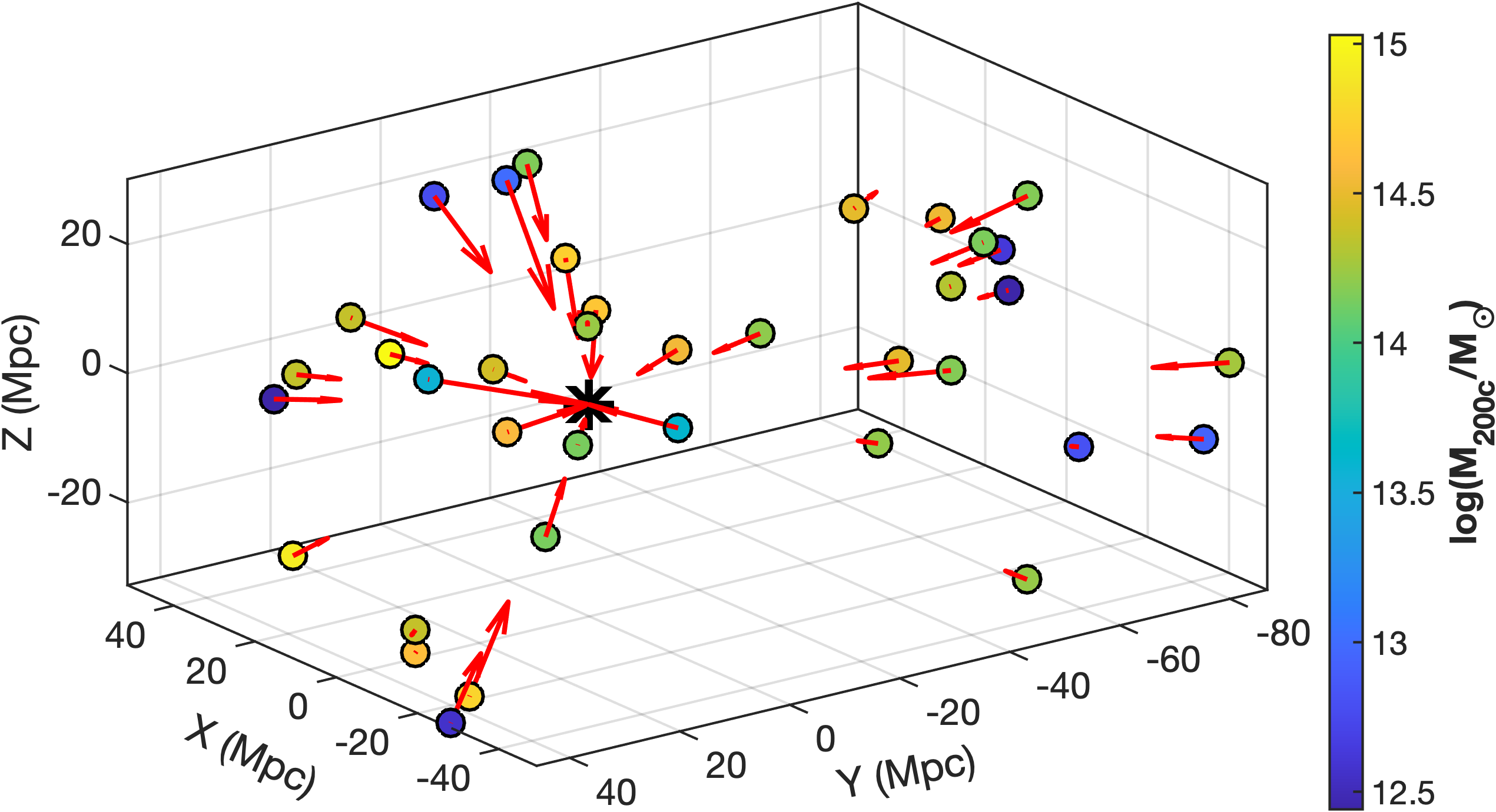}
    \caption{The most massive supercluster in the mock. The `$\ast$' symbol indicates the center of mass of the supercluster. Member clusters are represented by circles and their colors indicate the masses $\mathrm{M}_{200c}$. The red arrows show the supercluster-centric peculiar velocity components (radial peculiar velocities) of the member clusters with arrow lengths proportional to the radial peculiar velocities.
    Here, an arrow with a length of 1 Mpc corresponds to 60 km/s.
    }
    \label{fig:most_mass_pec_vel}
\end{figure*}

In Figure~\ref{fig:pec_vel}, we show the radial peculiar velocities as a function of their comoving distance from the centers of all the superclusters stacked in the reference frame of the center of mass of the supercluster. The radial peculiar velocity $v_{pec}$ of a member cluster is,
\begin{equation}
    v_{pec} = (\mathbf{v_{mem}} - \mathbf{v_{SC}}) \cdot \mathbf{\hat{r}}
\end{equation}
where $\mathbf{v_{mem}}$ is the peculiar velocity vector (as given in the HR4 simulation) of the member cluster, $\mathbf{v_{SC}}$ is the peculiar velocity vector of the center of mass of the supercluster and $\mathbf{\hat{r}}$ is the comoving position unit vector of the member cluster in the frame of reference of the center of mass of the supercluster.
Figure~\ref{fig:pec_vel} shows the phase space distribution of member clusters, where the y-axis is the radial peculiar velocities (peculiar velocity component of a member cluster in the reference frame of the center of mass and along the line from the center of mass to the member clusters) of member clusters of all superclusters and x-axis is the comoving distance of a member cluster from the center of mass of a supercluster. The black line represents a radial peculiar velocity of 0 km/s and the red line represents all the radial peculiar velocities equal and opposite to the Hubble flow as seen from the center of mass frame. The three regions divided by red and black lines tell us the kinematical state of the member clusters. \textit{Hubble Decoupled}: Member clusters that have decoupled from the Hubble flow and moving towards the supercluster center, \textit{Supercluster Influence}: Member clusters that have not decoupled from Hubble flow but have slowed down in their Hubble expansion and \textit{Outside Influence}: Member clusters that have their velocities greater than the Hubble flow and therefore have gravitational influence from regions outside of the supercluster. On average, 89\%, 9\%, and 2\% of the member clusters are in the regions of `Supercluster Influence', `Outside Influence', and `Hubble Decoupled', respectively.
For example, Figure~\ref{fig:most_mass_pec_vel} shows the supercluster-centric spatial distribution of the member clusters of the most massive supercluster in the mock. The arrows indicate the supercluster-centric peculiar velocity components (radial peculiar velocities), almost all pointing towards the supercluster center indicated by the `$\ast$' symbol.

\subsubsection{Phase Space Distribution of Mock Halos in and around Superclusters}
\begin{figure*}
    \centering
    \gridline{\fig{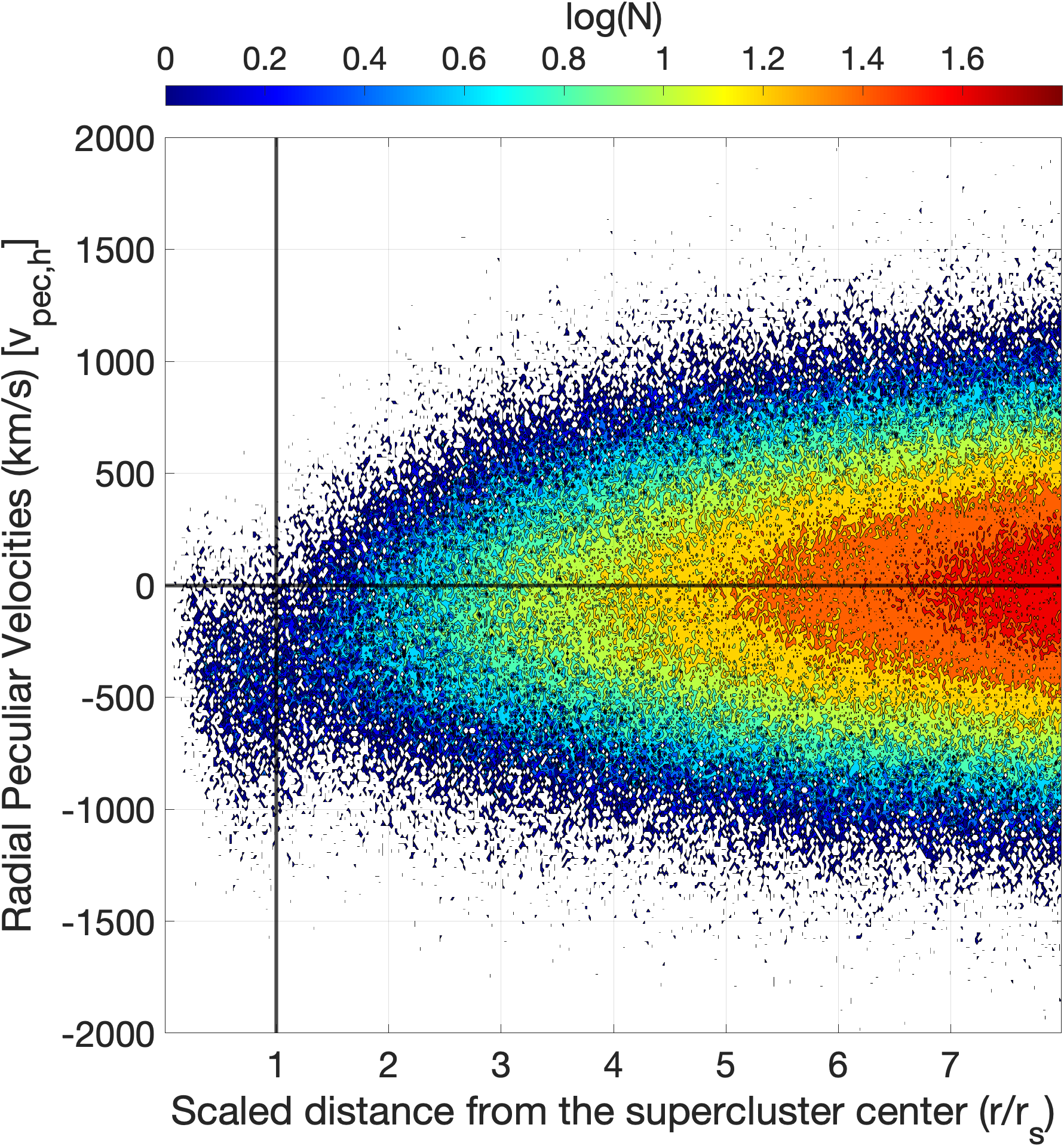}{0.46\textwidth}{(a)}
          \fig{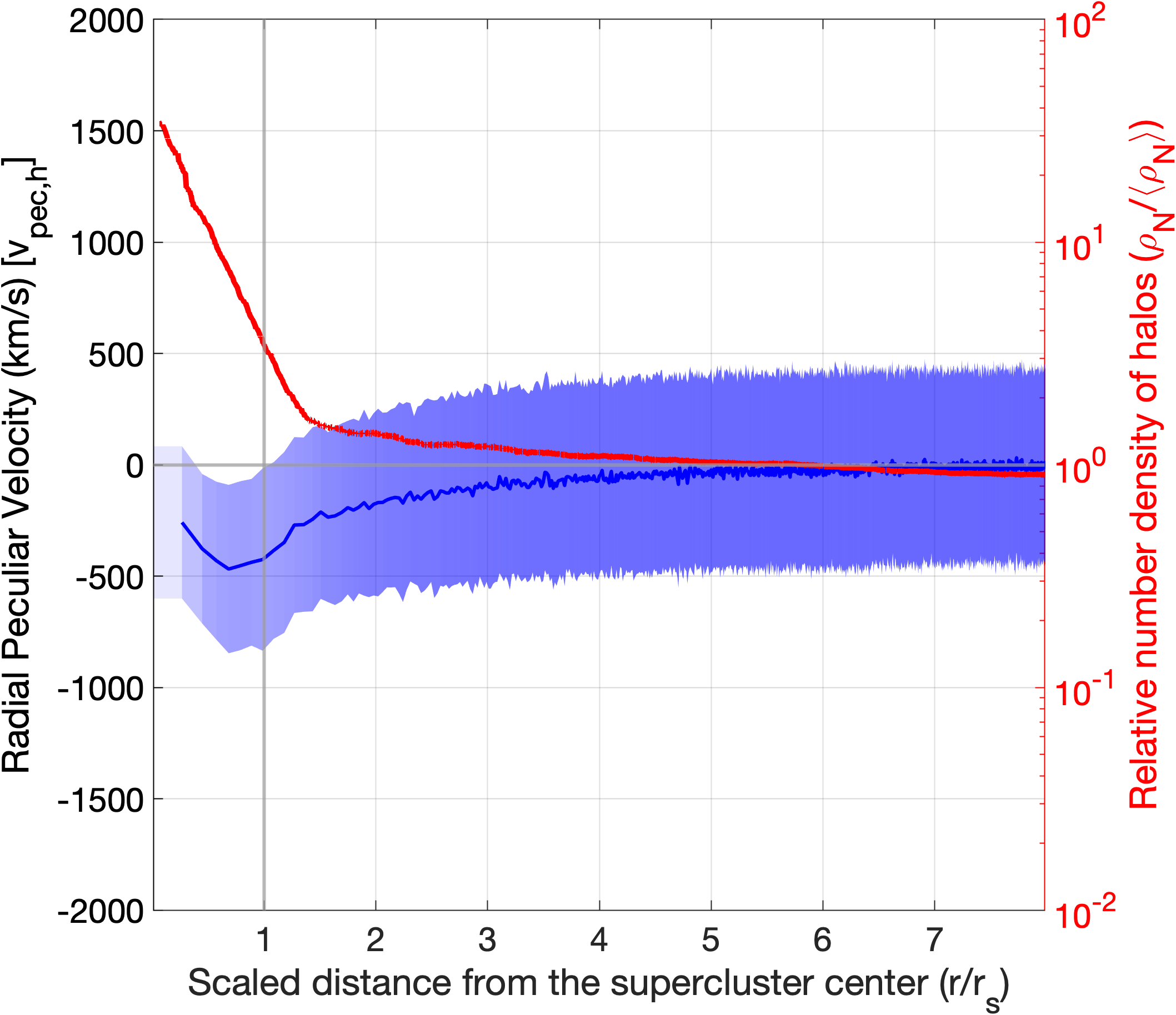}{0.515\textwidth}{(b)}}
    \caption{
    Stacked radial peculiar velocities and relative number density in and around superclusters. The x-axis denotes the distance from the supercluster center in units of $r_s$ (half the size of a supercluster).
    (a) Two-dimensional histogram of the mock halos in and around
    superclusters. $N$ denotes the number of halos in each grid. The region ($r/r_s \lesssim 1$ and $v_{pec,h} \lesssim 0$) contains most of the member halos of a supercluster. (b) Mean (blue line) values of the radial peculiar velocity component as a function of scaled distance. The shaded blue region shows the standard deviation (1 $\sigma$) of the radial peculiar velocity component. The red line shows the relative number density of halos as a function of scaled radius. This relative number density is calculated by dividing the average number density $\rho_N$ (of all stacked superclusters) at a scaled radius by the total number density $\langle\rho_N\rangle$ of halos within a sphere of radius $\sim 8 \times r_s$.
    }
    \label{fig:pec_vel_halos}
\end{figure*}

As shown in Figure~\ref{fig:pec_vel}, most supercluster member halos (91\%) have a negative supercluster-centric peculiar velocity component.
The member halos are influenced by the gravitational potential of their host supercluster. This influence should be negligible on the halos well outside the supercluster region.
The phase space distribution of halos within the supercluster region should be distinguishable from those outside the supercluster region. 
If we take the supercluster-centric peculiar velocity components of all halos from the supercluster center up to some distance outside the supercluster, we should see the overall negative components going to zero values at higher distances outside the supercluster. To explore this effect, we present the stacked supercluster-centric phase space distribution in and around the supercluster region in Figure~\ref{fig:pec_vel_halos}. The x-axis shows the scaled distance from the supercluster center.
If $S$ is the size of a supercluster, then we define a quantity, $r_s = S/2$, of a supercluster. The supercluster-centric comoving distances $r$ of halos are then scaled to $r_s$. This gives a unit-scaled distance as the rough extent of a supercluster. The left panel shows the two-dimensional histogram of the halos in and around the superclusters (stacked) up to a scaled distance, $r/r_s = 8$. In this panel, we can see that there is a detached distribution ( $r/r_s \lesssim 1$ and $v_{pec,h} \lesssim 0$) from the main distribution ($r/r_s \gtrsim 1$), where $v_{pec,h}$ is the supercluster-centric peculiar velocity component of halos. This detached distribution is the supercluster region where most halos have negative radial peculiar velocity components. Beyond the supercluster region ($r/r_s \gtrsim 1$), when it starts to encounter surrounding voids and other cosmic web components/structures, this distribution of the negative component of the peculiar velocity gradually and slowly moves towards zero.

The right panel of Figure~\ref{fig:pec_vel_halos} shows the mean values (blue line) of the radial peculiar velocities of halos at a certain scaled distance from the supercluster center. The shaded blue region shows the standard deviation (1 $\sigma$) of the supercluster-centric peculiar velocities distribution above and below the mean values. It shows that at a distance approximately 4--5 times $r_s$, the value goes to the global average of zero, which is expected on larger scales. The red line shows the relative number density of halos as a function of the scaled distance. It is calculated by dividing the average number density $\rho_N$ (of all stacked superclusters) at a scaled radius by the total number density $\langle\rho_N\rangle$ of halos within a sphere of radius $\sim 8 \times r_s$. Both velocity and density profiles show a characteristic scale at a distance of $\sim 1.5 \times r_s$.

It should be noted that the superclusters in our catalog are not spherically symmetric, and the supercluster center, which is the barycentre of the member clusters, may not coincide with the geometric center of the supercluster. Because of this asymmetry, the characteristic scale is not $r/r_s = 1$ and comes out to be $r/r_s \sim 1.5$. After this characteristic scale, both profiles gradually approach their global values.

Figure~\ref{fig:pec_vel} and Figure~\ref{fig:pec_vel_halos} support the robustness of the superclusters identified in our catalog. The density contrast of all the superclusters in our catalog has values greater than zero.
The relative density shows a knee around 1.5 times the scaled distance, also where the shape of the radial peculiar velocity profile changes. This suggests that these superclusters are well within the basin of attraction.
The overall densities of the superclusters are not sufficiently high enough to classify them as gravitationally bound structures within the framework of a spherical collapse model. Instead, they are more likely to fragment into smaller structures, similar to the observed cases of the Saraswati supercluster \citep{Bagchi17} and the BGW supercluster \citep{EinastoM22}.
We, therefore, conclude that our superclusters fall in the supercluster definition (2) -- the unbound over-dense regions in the Universe (see Section~\ref{sec:intro}).

\section{Discussion and Conclusion}
\label{sec:conc}
Superclusters, being the largest structures in the Universe, need more focused multi-wavelength studies than their current status of study. Whether they grow with a bottom-up process, top-down process, or both need to be addressed observationally as well as theoretically. Identifying superclusters and creating a statistically significant sample will increase our understanding of their properties and their environmental effects on the galaxies, groups, and clusters in them.

As the galaxy redshift surveys became deeper and larger over time, not only more superclusters but bigger superclusters have been discovered. Hence, it will not be surprising if even bigger superclusters are discovered in the upcoming redshift surveys like DESI \citep{Desi2016}, 4MOST \citep{de_Jong19,Richard19}, Euclid \citep{Laureijs11}, etc. And this may pose a challenge to the $\Lambda$CDM model of the Universe \citep{Park12}.

In this work, we identified 662 superclusters in the redshift range $0.05 \le z \le 0.42$ using the spatial distribution of WHL clusters. We applied a modified friends-of-friends algorithm to overcome the survey/catalog selection effects. These superclusters have typical masses $> 10^{15}$ M$_{\odot}$ and sizes $> 10$ Mpc.
\textit{Einasto Supercluster}, a new discovery at $z \sim 0.25$, is found to be the most massive supercluster in our catalog.

In our process of identifying superclusters, we rediscovered many previously known superclusters as well as discovered many new ones. We found a power law relation between a supercluster's density contrast and size with an index $\sim -2$.
Whether this suggests a relation with the morphology of the supercluster or some other relation is a matter of further study.
The topological analysis of the shape of superclusters may explain this phenomenon more clearly. Recently, \citet{Bag23} found that large superclusters with volumes $\gtrsim 10^4$ Mpc$^3$ tend to be more filamentary, and \citet{Heinamaki22} found that the low-luminosity, small, poor, and low-mass end of superclusters has pancake-type shapes while only a handful exhibit exceptionally spherical shapes.
The supercluster environment weakly affects the evolution of clusters. It is slightly more likely to find a massive cluster in a supercluster environment than in a non-supercluster environment.

The simulation gives slightly fewer high-mass clusters than those found in the observations. This gives a difference in the mass distribution of halos/clusters in the simulation and observational data and affects our comparison of the properties of superclusters found in the simulations and observations. Nevertheless, we found almost comparable properties in the simulated and SDSS data.
The peculiar velocity field of the mock superclusters in the simulation shows that most part of a supercluster points toward the center of mass of the supercluster. That is, most part of a supercluster is influenced by its mass and over-density. This leads to a slowing down of the expansion of the supercluster with respect to the Hubble flow.
The phase-space distribution of halos in and around the superclusters shows a characteristic length scale when a change in radial peculiar velocity and density profiles is seen. This length scale lies approximately 1.5 times the scaled distance from the supercluster center and arises due to the non-spherical shapes of the superclusters.

Superclusters host some of the most massive galaxy clusters formed through mergers. Under the right conditions, upon a merger, the resulting propagating shock manifests as Mpc scale radio relics \citep[e.g.][]{Bagchi02,Bagchi06,van_Weeren09} and radio halos \citep[for review see][]{vanWeeren19}.
Recently, using the eROSITA Final Equatorial Depth Survey, \citet{Ghirardini2021} reported a new supercluster at a redshift of $z \sim$\,0.36 with eight associated galaxy cluster members and spanning $\sim$\,27 Mpc. They have also carried out detailed x-ray, optical, and radio studies of the supercluster, where they found two new radio relics and a radio halo. Shapley Supercluster has been the subject of numerous multi-wavelength studies in the past few decades. Its central part has recently been studied at radio wavelengths using sensitive observations from uGMRT, MeerKAT, and ASKAP \citep{Venturi2022}. In addition to finding radio halos and very low surface brightness radio emissions connecting groups and clusters ($\sim$\,1 Mpc), such deep multi-frequency observations also enable to estimate of equipartition magnetic field ($\sim$\,0.76~$\muup$G) permeating across a large region. Hence, the study by \citet{Venturi2022} provides compelling evidence of non-thermal signatures detected from minor mergers (e.g., Mpc-scale radio emission from bridges connecting clusters and groups).

Simulations show that supercluster embryos form at very early cosmological epochs.
Even in observations, superclusters have been detected at higher redshifts, extending as far back as $z \sim 2.45$ \citep{Lietzen16, Kim2016, Cucciati2018, Shimakawa2023}.
The location of superclusters does not change much
during evolution, and the essential evolutionary changes occur within the supercluster
cocoons (basins of attraction or the regions of dynamical influence) \citep{Einasto19,Einasto21}. On the question of the formation of the superclusters, \citet{Einasto11} and \citet{Sukhonenko11} using wavelet analysis of the cosmic web, show that the rich clusters and superclusters form where density waves of medium ($\simeq 32$ h$^{-1}$Mpc) and large scales ($\geq 64$ h$^{-1}$Mpc) combine in similar phases to generate high-density regions. They show that the synchronization (or coupling) of different scaled density waves plays an important role in structure formation. Their results show that the largest structures to form have sizes of approximately 100 h$^{-1}$Mpc, which agrees with the size of the largest superclusters in our study. Density  waves with larger scales only modify the properties of structures.

Under-dense regions (voids) which surround superclusters form where different phases of the density waves are combined. Matter within the voids flows toward the inner regions of superclusters
due to gravitational instability. Observationally, this has been shown by analyzing local velocity fields \citep{Tully14,Hoffman2017}. The voids may be expanding more rapidly than the global expansion rate and thus helping in the growth of the supercluster \citep{Sheth04, Waygaert16, Hoffman2017, Bagchi17}. The expansion of large voids can facilitate shaping surroundings and help in the formation of long coherent structures (superclusters) along its boundary or periphery.

Superclusters are important to study the formation of structures on large scales. For further advances in this subject, a detailed spectroscopic survey of a massive supercluster is needed (e.g., Saraswati supercluster), which can inform more about the dynamical state of the supercluster. Such surveys will provide crucial information needed to answer open questions in Supercluster physics, provide a wealth of data for individual galaxies, and pave the path for more synergistic studies.
Future deep, wide sky galaxy surveys will provide the opportunity to compare the abundance and properties of superclusters at higher redshifts with the ones in simulations. Therefore, it is vital to identify and characterize superclusters. It is important to compare the properties of galaxies in large, high-density regions like superclusters and in under-dense regions like voids to understand the factors affecting their growth and evolution. The current era of deep multi-wavelength large sky surveys provides us with the perfect opportunity.

\newpage
\section*{Acknowledgements}
We thank the referee for the valuable and constructive comments, which have enhanced the paper's clarity and readability.
SS acknowledges the support of the European Regional Development Fund, the Mobilitas Pluss postdoctoral research grant MOBJD660, and the ETAg grant PRG1006. ET and ME acknowledge the support by ETAg grant PRG1006 and EU through the ERDF CoE TK133. JB acknowledges the help and full support from Christ (Deemed to be) University in facilitating the research. We kindly thank Changbom Park, Sung-Wook Kim, and Juhan Kim for providing the HR4 simulation products on request.

\bibliography{Sankhyayan_superclusters}

\appendix
\counterwithin{figure}{section}
\section{Five most massive superclusters}
\label{sec:top5}
\begin{figure*}[htp]
\gridline{\fig{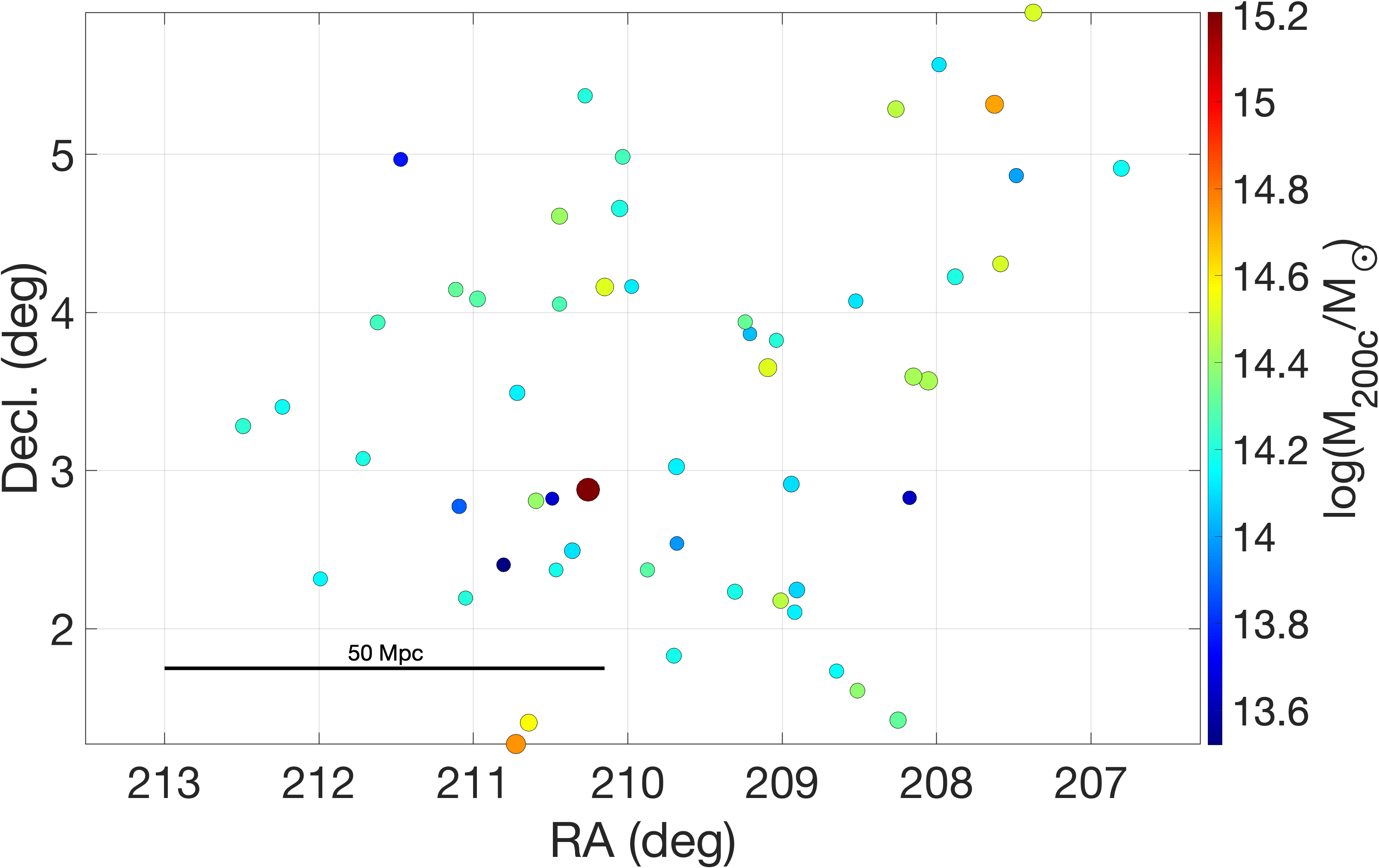}{0.45\textwidth}{SCl 1}
          \fig{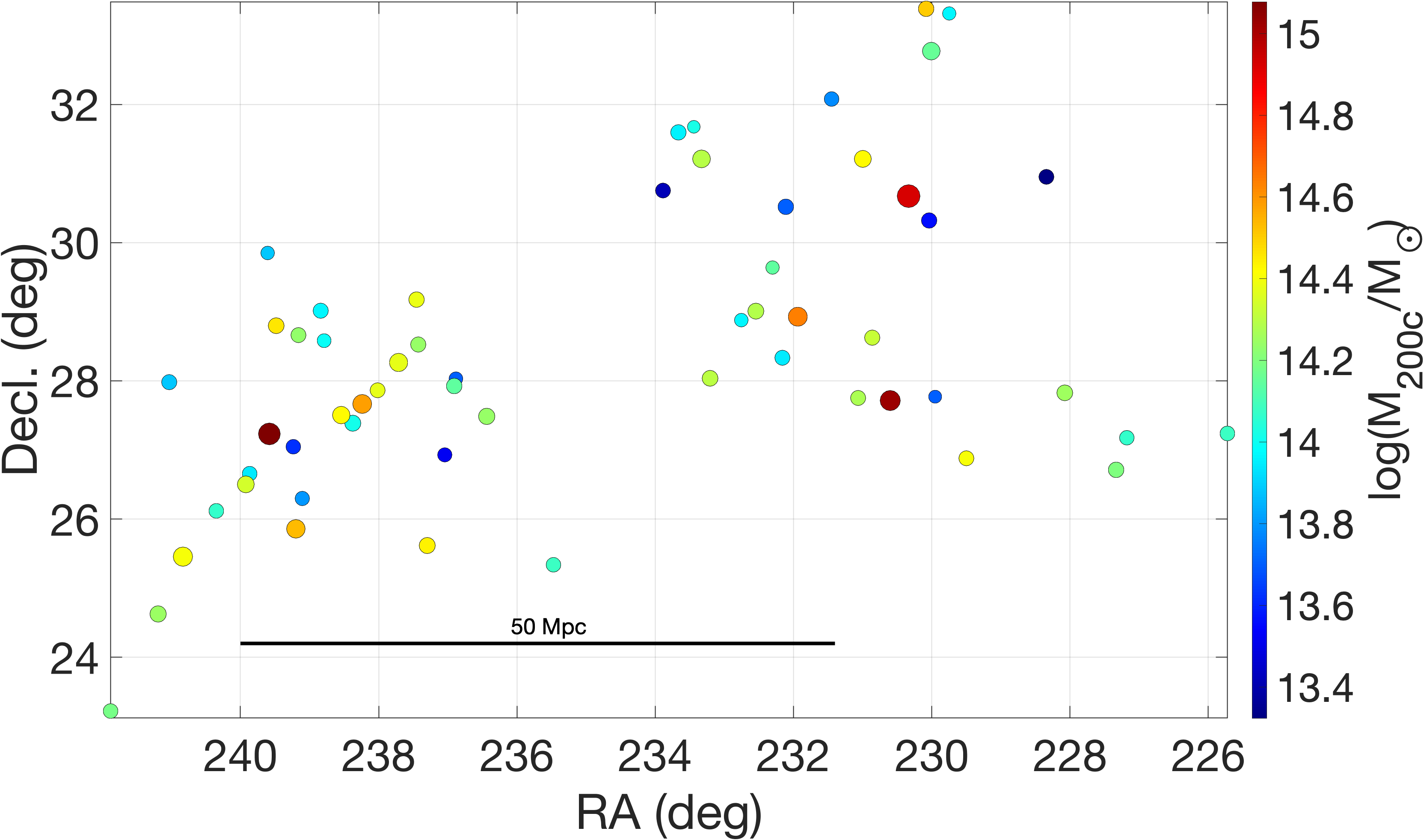}{0.47\textwidth}{SCl 2 : Corona Borealis}
          }
\gridline{\fig{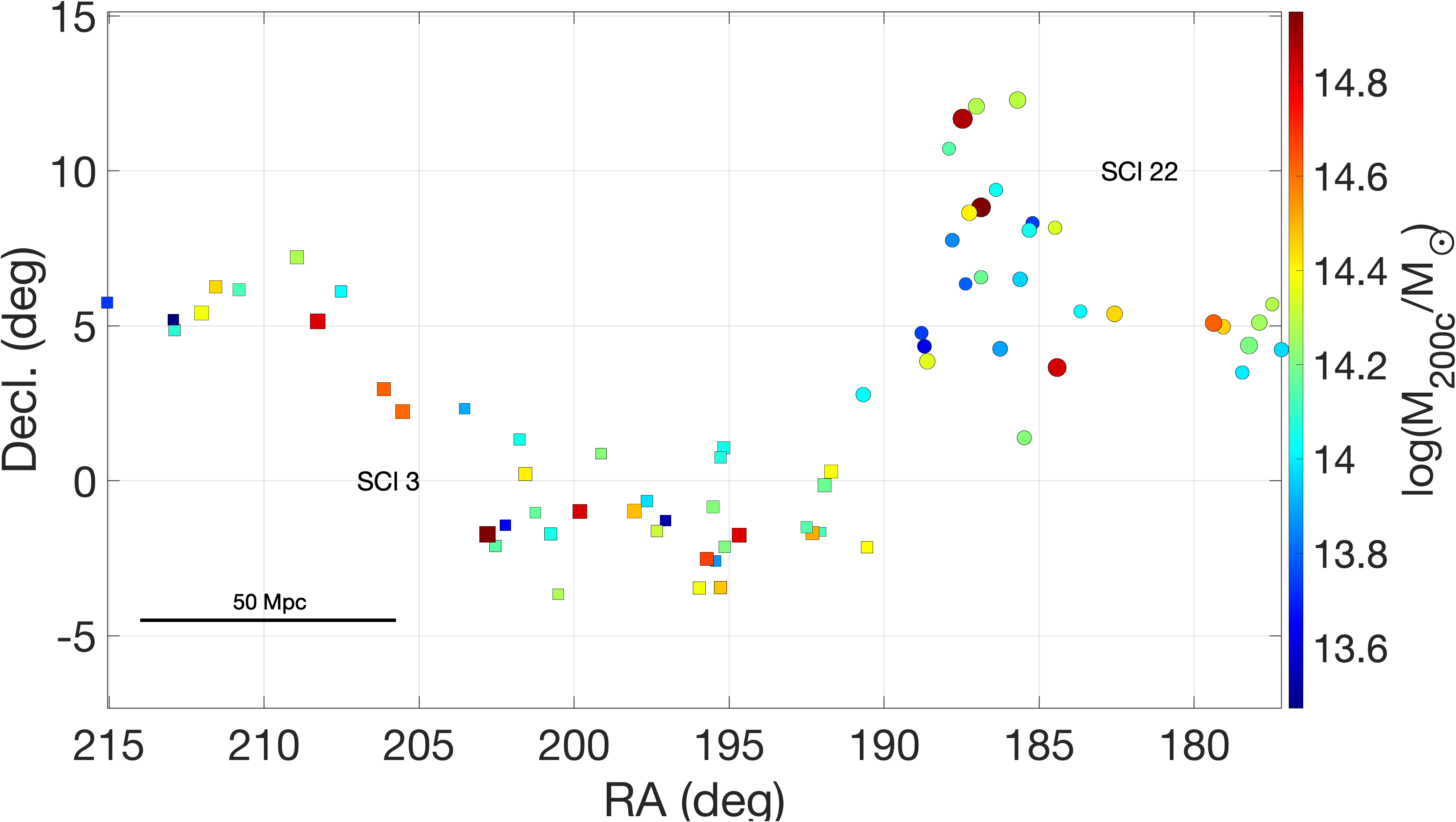}{0.50\textwidth}{Sloan Great Wall}
          \fig{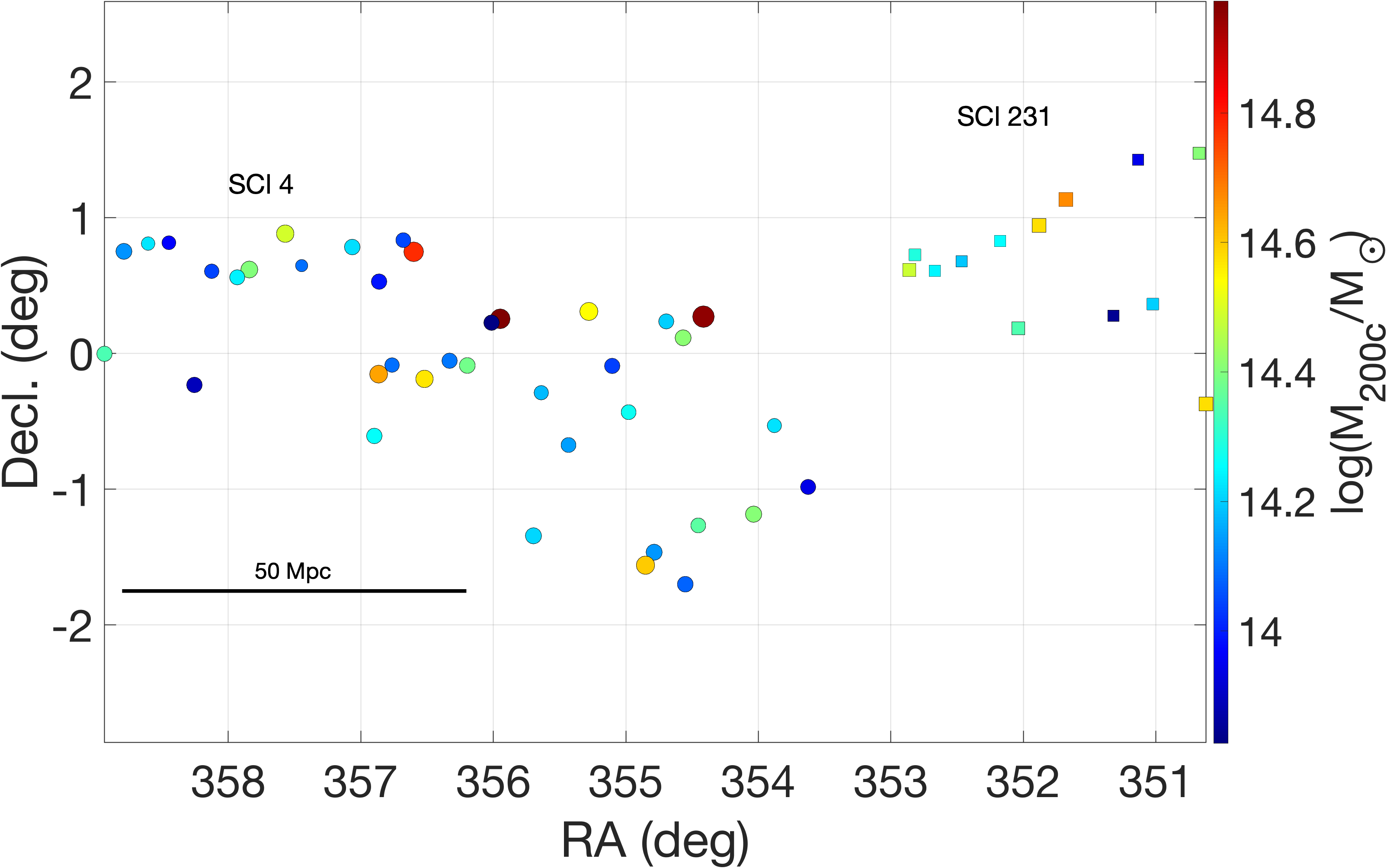}{0.45\textwidth}{Saraswati}
          }
\gridline{\fig{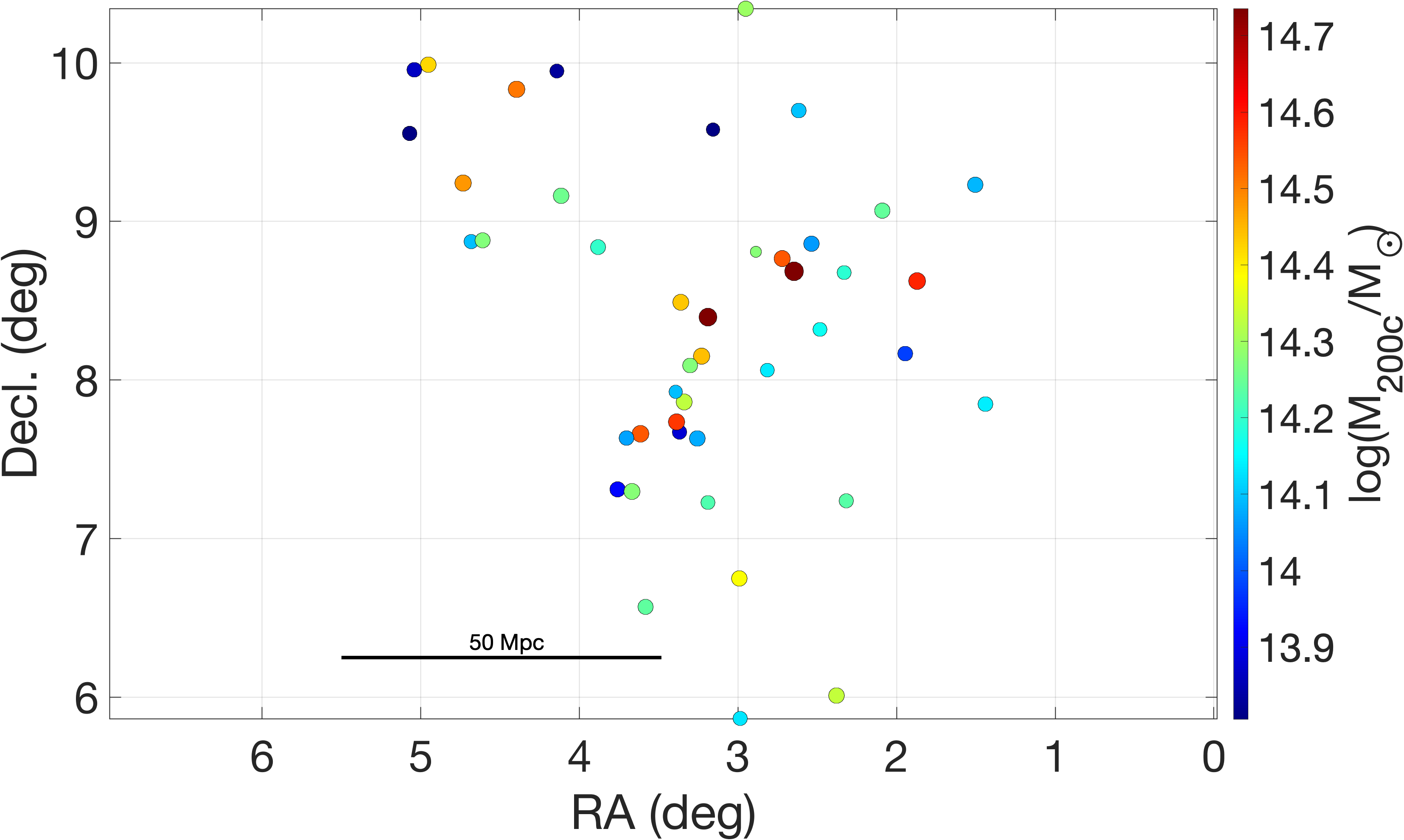}{0.45\textwidth}{SCl 5}}
\caption{
Sky plane distribution of the five most massive superclusters identified in our supercluster catalog.
}
\label{fig:scl1-5}
\end{figure*}

\begin{figure*}
\gridline{\fig{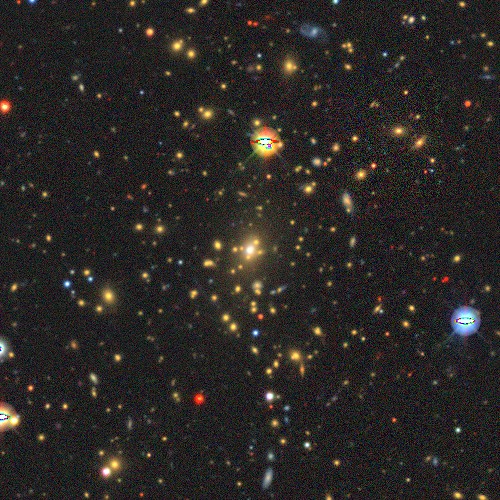}{0.32\textwidth}{SCl 1 (4\arcmin.2 x 4\arcmin.2) \\ Abell 1835 : RA = 210\arcdeg.25862, Decl. = 2\arcdeg.87847}
          \fig{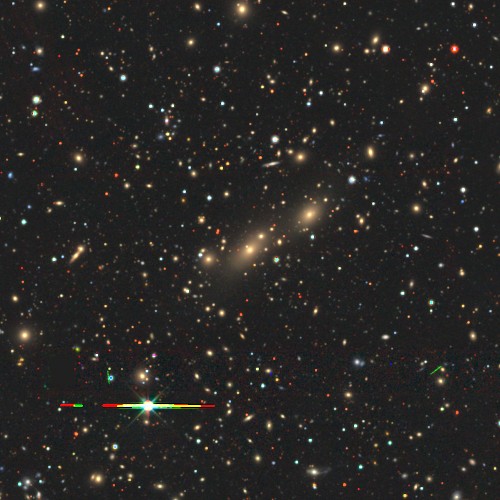}{0.32\textwidth}{SCl 2 (12\arcmin.5 x 12\arcmin.5) \\ Abell 2142 : RA = 239\arcdeg.58334, Decl. = 27\arcdeg.23341}}
\gridline{\fig{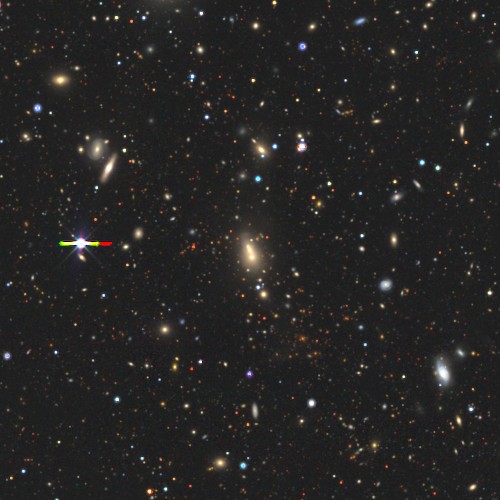}{0.32\textwidth}{SCl 3 (12\arcmin.5 x 12\arcmin.5) \\ Abell 1750N : RA = 202\arcdeg.79594, Decl. = -1\arcdeg.72730}
          \fig{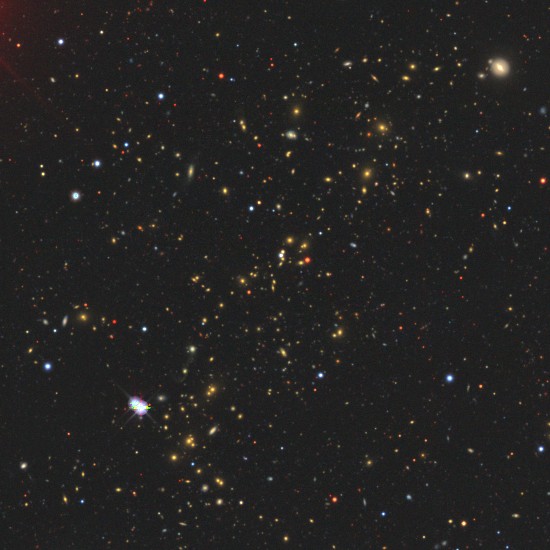}{0.32\textwidth}{SCl 4 (9\arcmin.2 x 9\arcmin.2) \\ ZwCl 2341+0000 : RA = 355\arcdeg.94806, Decl. = 0\arcdeg.25666}}
\gridline{\fig{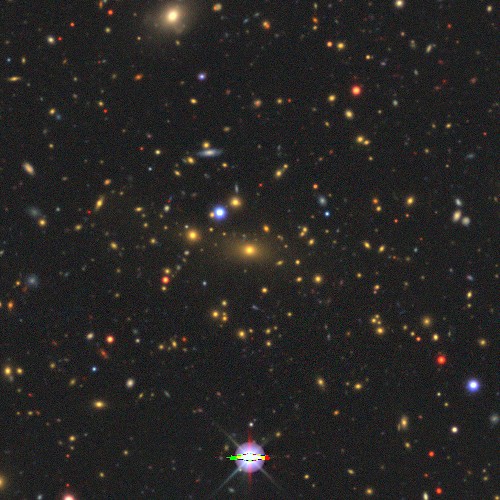}{0.32\textwidth}{Abell 2631 (4\arcmin.2 x 4\arcmin.2) \\ RA = 354\arcdeg.41554, Decl. = 0\arcdeg.27137}
          \fig{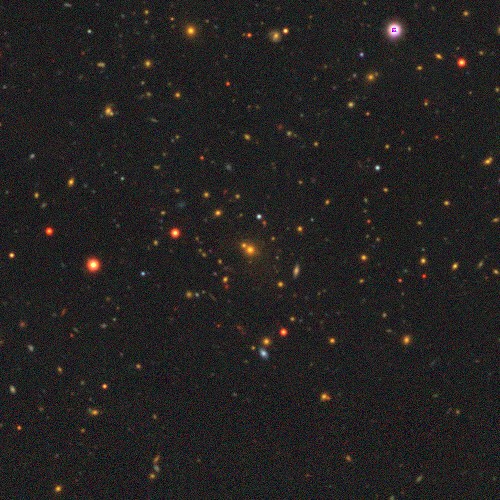}{0.32\textwidth}{SCl 5 (4\arcmin.2 x 4\arcmin.2) \\ RA = 3\arcdeg.18930, Decl. = 8\arcdeg.39641}}
\caption{Optical images of the most massive member clusters in each of the top five superclusters in Table~\ref{tab:WHLsupclustcat}. Abell 2631 is actually the most massive cluster of SCl 4 (Saraswati supercluster). The images are extracted from DESI Legacy Imaging Surveys (DR9).}
\label{fig:optical_images}
\end{figure*}

Figure~\ref{fig:scl1-5} shows the sky plane distribution of the five most massive superclusters. The member clusters are shown with their sizes proportional to their R$_{200c}$, and the color scale indicates their masses (M$_{200c}$). Different marker types denote different superclusters.
The black horizontal line shows a comoving scale of 50 Mpc.
Figure~\ref{fig:optical_images} shows the optical images of
the most massive member clusters of the top five superclusters in Table~\ref{tab:WHLsupclustcat}. For more details, see section~\ref{sec:mast_mass_sc}.

\end{document}